\documentclass{amsart}
\def\Defined{}
\ifx\UseRussian\Defined
	\usepackage[T2A,T2B]{fontenc}
	\usepackage[cp1251]{inputenc}
	\usepackage[english,russian]{babel}
\fi
\usepackage{footmisc}
\usepackage[all]{xy}
\usepackage{color}
\definecolor{UrlColor}{rgb}{.9,0,.3}
\definecolor{SymbColor}{rgb}{.4,0,.9}
\definecolor{IndexColor}{rgb}{1,.3,.6}
\definecolor{eml1}{rgb}{.8,.1,.1}
\definecolor{eml2}{rgb}{.1,.6,.6}

\usepackage{xr-hyper}
\usepackage[unicode]{hyperref}
\hypersetup{pdfdisplaydoctitle=true}
\hypersetup{colorlinks}
\hypersetup{citecolor=UrlColor}
\hypersetup{urlcolor=UrlColor}
\hypersetup{pdffitwindow=true}
\hypersetup{pdfnewwindow=true}
\hypersetup{pdfstartview={FitH}}

\def\hyph{\penalty0\hskip0pt\relax-\penalty0\hskip0pt\relax}
\def\Hyph{-\penalty0\hskip0pt\relax}

\newcommand{\Basis}[1]{\overline{\overline{#1}}{}}
\newcommand{\Vector}[1]{\overline{#1}{}}
\newcommand{\gi}[1]{\boldsymbol{\textcolor{IndexColor}{#1}}}
\makeatletter
\newcommand{\NameDef}[1]{%
	\expandafter\gdef\csname #1\endcsname%
}%
\newcommand{\ShowSymbol}[1]{%
	\@nameuse{ViewSymbol#1}%
}%
\newcommand{\symb}[3]{%
	\@ifundefined{ViewSymbol#3}{%
		\NameDef{ViewSymbol#3}{\textcolor{SymbColor}{#1}}%
		\NameDef{RefSymbol#3}{\pageref{symbol: #3}}%
		\@namedef{LabelSymbol#3}{\label{symbol: #3}}%
	}{%
		\NameDef{RefSymbol#3}{}%
		\@namedef{LabelSymbol#3}{}%
	}%
	\ifcase#2
	\or
		$\@nameuse{ViewSymbol#3}$%
	\or
		\[\@nameuse{ViewSymbol#3}\]%
	\else%
	\fi%
	\@nameuse{LabelSymbol#3}%
}%
\makeatother

\newcommand{\subs}{${}_*$\Hyph}
\newcommand{\sups}{${}^*$\Hyph}

\newcommand{\CRstar}{{}_*{}^*}
\newcommand{\RCstar}{{}^*{}_*}

\newcommand{\RC}{$\RCstar$\Hyph}
\newcommand{\CR}{$\CRstar$\Hyph}
\newcommand{\drc}{$D\RCstar$\Hyph}
\newcommand{\dcr}{$D\CRstar$\hyph}
\newcommand{\rcd}{$\RCstar D$\Hyph}
\newcommand{\crd}{$\CRstar D$\Hyph}

\newcommand\sT{$\star T$\Hyph}%
\newcommand\Ts{$T\star$\Hyph}%

\renewcommand{\uppercasenonmath}[1]{}

\makeatletter
\newcommand\@dotsep{4.5}
\def\@tocline#1#2#3#4#5#6#7
{\relax
		\par \addpenalty\@secpenalty\addvspace{#2}%
		\begingroup \hyphenpenalty\@M
		\@ifempty{#4}{%
			\@tempdima\csname r@tocindent\number#1\endcsname\relax
		}{%
			\@tempdima#4\relax
		}%
		\parindent\z@ \leftskip#3\relax \advance\leftskip\@tempdima\relax
		\rightskip\@pnumwidth plus1em \parfillskip-\@pnumwidth
		#5\leavevmode\hskip-\@tempdima #6\relax
		\leaders\hbox{$\m@th
		\mkern \@dotsep mu\hbox{.}\mkern \@dotsep mu$}\hfill
		\hbox to\@pnumwidth{\@tocpagenum{#7}}\par
		\nobreak
		\endgroup
}
\makeatother 

\ifx\PrintBook\undefined
	\usepackage{fancyhdr}
	\makeatletter
	\renewcommand{\@indextitlestyle}{%
		\twocolumn[\section{\indexname}]%
		\def\IndexSpace{off}%
	}
	\makeatother 
	\thanks{\href{mailto:Aleks\_Kleyn@MailAPS.org}{Aleks\_Kleyn@MailAPS.org}}
\else
	\makeatletter
	\renewcommand{\@indextitlestyle}{%
		\twocolumn[\chapter{\indexname}]%
		\def\IndexSpace{off}%
		\let\@secnumber\@empty
		\chaptermark{\indexname}%
	}
	\makeatother 
	\email{\href{mailto:Aleks\_Kleyn@MailAPS.org}{Aleks\_Kleyn@MailAPS.org}}
\fi

\ifx\SelectlEnglish\undefined
	\ifx\UseRussian\undefined
		\def\SelectlEnglish{}
	\fi
\fi

\ifx\SelectlEnglish\undefined
	\newcommand\CurrentLanguage{Russian.}%
	\author{Александр Клейн}
	\newtheorem{theorem}{Теорема}[section]
	\newtheorem{corollary}[theorem]{Следствие}
	\theoremstyle{definition}
	\newtheorem{definition}[theorem]{Определение}
	\newtheorem{example}[theorem]{Пример}
	\newtheorem{xca}[theorem]{Exercise}
	\theoremstyle{remark}
	\newtheorem{remark}[theorem]{Замечание}
	
	\newcommand\Gbasis{$G$\Hyph базис}
	\newcommand\Gcoords{$G$\Hyph координат}
	\newcommand\Gspace{$G$\Hyph пространств}
	\newcommand\xRefDef[2]
		{
		\externaldocument[#1-Russian-]{#1.Russian}[http://arxiv.org/PS_cache/#2.pdf]
		\NameDef{xRefDef#1}{}%
		}
	\makeatletter
	\newcommand\xRef[2]%
	{%
		\@ifundefined{xRefDef#1}{%
		\ref{#2}%
		}{%
		\citeBib{#1}-\ref{#1-Russian-#2}%
		}%
	}%
	\newcommand\xEqRef[2]%
	{%
		\@ifundefined{xRefDef#1}{%
		\eqref{#2}%
		}{%
		\citeBib{#1}-\eqref{#1-Russian-#2}%
		}%
	}%
	\makeatother
	\ifx\PrintBook\undefined
		\newcommand{\BibTitle}{%
			\section{Список литературы}%
		}
	\else
		\newcommand{\BibTitle}{%
			\chapter{Список литературы}%
		}
	\fi
\else
	\newcommand\CurrentLanguage{English.}%
	\author{Aleks Kleyn}
	\newtheorem{theorem}{Theorem}[section]
	\newtheorem{corollary}[theorem]{Corollary}
	\theoremstyle{definition}
	\newtheorem{definition}[theorem]{Definition}

	\theoremstyle{remark}
	
	\newcommand\Gbasis{$G$\Hyph basis}
	\newcommand\Gcoords{$G$\Hyph coordinates}
	\newcommand\Gspace{$G$\Hyph space}
	\newcommand\xRefDef[2]
		{
		\externaldocument[#1-English-]{#1.English}[http://arxiv.org/PS_cache/#2.pdf]
		\NameDef{xRefDef#1}{}%
		}
	\makeatletter
	\newcommand\xRef[2]%
	{%
		\@ifundefined{xRefDef#1}{%
		\ref{#2}%
		}{%
		\citeBib{#1}-\ref{#1-English-#2}%
		}%
	}%
	\newcommand\xEqRef[2]%
	{%
		\@ifundefined{xRefDef#1}{%
		\eqref{#2}%
		}{%
		\citeBib{#1}-\eqref{#1-English-#2}%
		}%
	}%
	\makeatother
	\ifx\PrintBook\undefined
		\newcommand{\BibTitle}{%
			\section{References}%
		}
	\else
		\newcommand{\BibTitle}{%
			\chapter{References}%
		}
	\fi
\fi

\ifx\PrintBook\undefined
	\numberwithin{Hfootnote}{section}
\else
	\numberwithin{section}{chapter}
	\numberwithin{footnote}{chapter}
	\numberwithin{Hfootnote}{chapter}
\fi

\numberwithin{equation}{section}
\numberwithin{figure}{section}
\numberwithin{table}{section}
\numberwithin{Item}{section}

\makeatletter
\newcommand\org@maketitle{}
\let\org@maketitle\maketitle
\def\maketitle{%
	\hypersetup{pdftitle={\@title}}%
	\hypersetup{pdfauthor={\authors}}%
	\hypersetup{pdfsubject=\@keywords}%
	\org@maketitle
}
\def\make@stripped@name#1{%
	\begingroup
		\escapechar\m@ne
		\global\let\newname\@empty
		\protected@edef\Hy@tempa{\CurrentLanguage #1}%
		\edef\@tempb{%
			\noexpand\@tfor\noexpand\Hy@tempa:=%
			\expandafter\strip@prefix\meaning\Hy@tempa
		}%
		\@tempb\do{%
			\if\Hy@tempa\else
				\if\Hy@tempa\else
					\xdef\newname{\newname\Hy@tempa}%
				\fi
			\fi
		}%
	\endgroup
}%
\newenvironment{enumBib}{%
	\BibTitle
	\advance\@enumdepth \@ne
	\edef\@enumctr{enum\romannumeral\the\@enumdepth}\list
	{\csname biblabel\@enumctr\endcsname}{\usecounter
	{\@enumctr}\def\makelabel##1{\hss\llap{\upshape##1}}}
}{%
	\endlist
}

\def\Chapters#1{\ChapterList#1,LastChapter,}%
\def\LastChapter{LastChapter}%
\def\ChapterList#1,{\def\temp{#1}%
	\ifx\temp\LastChapter
	\else
		\@ifundefined{#1}{%
		}{%
			\def\Semafor{on}
		}
		\expandafter\ChapterList
	\fi
}%
\newcommand{\BiblioItem}[3]
{
	\def\Semafor{off}
	\Chapters{#1}
	\ifx\Semafor\ValueOn
		\ifx\IndexState\ValueOff
			\begin{enumBib}
			\def\IndexState{on}
		\fi
		\item \label{bibitem: #2}#3%
	\fi
}
\newcommand{\OpenBiblio}
{
	\def\IndexState{off}
}
\newcommand{\CloseBiblio}
{
	\ifx\IndexState\ValueOn
		\end{enumBib}
		\def\IndexState{off}
	\fi
}

\def\StartCite{[}%
\def\citeBib#1{\protect\showCiteBib#1,endCite,}%
\def\endCite{endCite}%
\def\showCiteBib#1,{\def\temp{#1}%
\ifx\temp\endCite
]%
\def\StartCite{[}%
\else
	\StartCite\ref{bibitem: #1}%
	\def\StartCite{, }%
\expandafter\showCiteBib%
\fi}%

\makeatother 
\newcommand{\arp}{\ar @{-->}}

\newcommand{\bundle}[4]%
{%
	\def\tempa{}%
	\def\tempb{#3}%
	\def\tempc{#1}%
	\ifx\tempa\tempb%
		\ifx\tempa\tempc%
			#2%
		\else%
			\xymatrix{#2:#1\arp[r]&#4}%
		\fi%
	\else%
		\ifx\tempa\tempc%
			#2[#3]%
		\else%
			\xymatrix{#2[#3]:#1\arp[r]&#4}%
		\fi%
	\fi%
}%
\newcommand{\AddIndex}[2]%
{%
	{\bf #1}%
	\label{index: #2}%
}%
\newcommand{\Index}[3]%
{%
	\def\Semafor{off}%
	\Chapters{#1}%
	\ifx\Semafor\ValueOn%
		\def\tempa{}%
		\def\tempb{#3}%
		\ifx\IndexState\ValueOff%
			\begin{theindex}%
			\def\IndexState{on}%
		\fi%
		\ifx\IndexSpace\ValueOn%
			\indexspace%
			\def\IndexSpace{off}%
		\fi%
		\item #2%
		\ifx\tempa\tempb%
		\else%
			\ \pageref{index: #3}%
		\fi%
	\fi%
}%
\newcommand{\SubIndex}[3]
{
	\def\Semafor{off}
	\Chapters{#1}
	\ifx\Semafor\ValueOn
		\subitem #2 \pageref{index: #3}
	\fi
}%

\makeatletter
\newcommand{\Symb}[3]
{
	\def\Semafor{off}
	\Chapters{#1}
	\ifx\Semafor\ValueOn
		\ifx\IndexState\ValueOff
			\begin{theindex}
			\def\IndexState{on}
		\fi
		\ifx\IndexSpace\ValueOn
			\indexspace
			\def\IndexSpace{off}
		\fi
		\item $\@nameuse{ViewSymbol#3}$\ \ #2
		\@nameuse{RefSymbol#3}%
	\fi
}

\makeatother

\newcommand{\SetIndexSpace}%
{%
	\def\IndexSpace{on}%
}%
\def\ValueOff{off}
\def\ValueOn{on}

\newcommand{\OpenIndex}
{
	\def\IndexState{off}
}
\newcommand{\CloseIndex}
{
	\ifx\IndexState\ValueOn
		\end{theindex}
		\def\IndexState{off}
	\fi
}

\def\LastMemo{LastMemo}%
\def\MemoList#1//{\def\temp{#1}%
	\ifx\temp\LastMemo
	\else%
		\par\setlength{\parindent}{12pt}\textcolor{blue}{#1}%
		\expandafter\MemoList%
	\fi%
}%

\def\texPrefaceTorsion{}
\xRefDef{0405.027}{gr-qc/pdf/0405/0405027v3}

\begin{document}
\title{Metric-Affine Manifold}

\thanks{I wish to express my thanks to my teachers Gavrilchenko Michael Leonidovich and
Rahula Meido Oscarovich for their help and support in my research.}

\keywords{Differential geometry, general relativity, metric-affine manifold,
Frenet transport, extreme line, Killing equation, Lie derivative}

\pdfbookmark[1]{Metric-Affine Manifold}{TitleEnglish}
\begin{abstract}
We call a manifold with torsion and nonmetricity the metric\hyph affine manifold.
The nonmetricity leads to a difference between the auto parallel line and the extreme line,
and to a change in the expression of the Frenet transport and moving basis.
The torsion leads to a change in the Killing equation. We also need to add a similar equation
for the connection.

The analysis of the Frenet transport leads to the concept of the Cartan transport
and an introduction of the connection compatible with the metric tensor.
The dynamics of a particle follows to the Cartan transport.
We need additional physical constraints to make
a nonmetricity observable.
\end{abstract}

\maketitle

\def\texIntro{}

\ifx\PrintBook\Defined
				\chapter{Introduction}
			\section{About This Book}
			\label{section: About This Book}

Sometimes is very hard to give name for book that you want to write.
Even you wrote this book the whole life. And may be for this reason.
The way is not finished. However I feel it is time to share with people
my discoveries. Only future will tell which parts of this research
will be useful.
This book starts from learning of geometric object, turns to
reference frame in physics, then suddenly changes its course
to learn metric affine manifolds.

Since Einstein introduced general relativity, the close relation between geometry
and physics became a reality. At the same time, quantum mechanics introduced new concepts that
contradicted a tradition established during centuries.
This meant that we need new geometrical concepts that will become part
of the language of quantum mechanics.
This is the reason for returning to the beginning.

The whole my life was dedicated to solve one of the greatest
mysteries that I met in the beginning of my life. When I learned
general relativity and quantum mechanics I felt that language of
quantum mechanics is not adequate to phenomenon that it observes.
I mean the geometry.

I dedicated chapter \ref{chapter: Space and Time in Physics}
to small essay that I wrote when I was young.\footnote{Unfortunately
some references are lost. If somebody recognize
\label{footnote: Space and Time in Physics}
familiar text, I will appreciate if he let me know exactly reference.}
\fi

\ifx\texPrefaceRefernceFrame\Defined
\section{Geometrical Object and Invariance Principle}

\ifx\PrintBook\Defined
Sections \ref{section: Basis in Vector Space}
and \ref{section: Geometrical Object of Vector Space}
was written under the great influence of the book \citeBib{Rashevsky}.
The studying of a homogenous space of a group of symmetry of a vector space
leads us to the definition of a basis of this space. A basis  manifold is a set of
bases of particular vector space and is an example of a homogenous space.
As it is shown in \citeBib{Rashevsky} it gives ability to define
concepts of invariance and of geometrical object.

We introduce two types of transformation of a basis manifold:
active and passive transformations. The difference between them is
that the passive transformation can be expressed as a transformation of
an original space.

This definition can be extended on an arbitrary manifold. However in
this case we generalize the definition of a basis and introduce a reference
frame. In case of an event space of general relativity it leads us to
a natural definition of a reference frame and the Lorentz transformation.
A reference frame in event space is a smooth field of orthonormal basis.
\fi

The invariance principle which we studied \citeBib{0412.391}
is limited by vector spaces and we can use it only
in frame of the special relativity. Our task is
to describe structures which allow to extend
the invariance principle to general relativity.

A measurement of a spatial interval and a time length is one of the major tasks of general relativity.
This is a physical process that allows the study of geometry in a certain area of spacetime.
From a geometrical point of view,
the observer uses an orthonormal basis in a tangent plane as his measurement tool
because an orthonormal basis leads to the simplest local geometry.
When the observer moves from point to point he brings his measurement tool with him.

Notion of a geometrical object is closely related with
physical values measured in space time.
The invariance principle allows expressing physical laws
independently from the selection of a basis.
On the other hand if we want
to examine a relationship obtained in the test, we need to select
measurement tool. In our case this is basis. Choosing
the basis we can define coordinates of the geometrical object
corresponding to studied physical value.
Hence we can define the measured value.

Every reference frame is equipped by anholonomic coordinates.
For instance, synchronization of a reference frame is an anholonomic time coordinate.
Simple calculations show how synchronization influences time measurement in
the vicinity of the Earth.
Measurement of Doppler shift from the star orbiting the black hole helps
to determine mass of the black hole.

Sections \ref{section: Time Delay in Central Body Gravitational Field}
and \ref{section: Lorentz Transformation in Orbital Direction}
show importance of calculations in orthogonal
frame. Coordinates that we use in event space are just labels
and calculations that we make in coordinates may appear
not reliable. For instance in papers \citeBib{Tartaglia,Tomozawa}
authors determine coordinate speed of light. This leads to
not reliable answer and as consequence of this to
the difference of speed of light in different directions.
\fi

\ifx\texPrefaceTidal\Defined
	\def\texPrefaceTidalOrPrefaceRefernceFrame{}
\fi
\ifx\texPrefaceRefernceFrame\Defined
	\def\texPrefaceTidalOrPrefaceRefernceFrame{}
\fi
\ifx\texPrefaceTidalOrPrefaceRefernceFrame\Defined
Paper \citeBib{Ranada} drew my attention. To explain anomalous acceleration of Pioneer 10
and Pioneer 11 (\citeBib{Anderson02}) Antonio Ranada incorporated the old Einstein's view on nature of
gravitational field and considered Einstein's idea about variability of speed of light.
When Einstein started to study the gravitational field he tried to
keep the Minkowski geometry, therefore he assumed that scale of
space and time does not change. As result he had to accept the idea that speed of light should
vary in gravitational field. When Grossman introduced Riemann geometry to Einstein, Einstein
realized that the initial idea was wrong and Riemann geometry solves his problem better.
Einstein never returned to idea about variable speed of
light.

Indeed, three values: scale of length and time and speed of light are correlated in present
theory and we cannot change one without changing another.
The presence of gravitational field changes this relation. We have two choices. We keep
a priory given geometry (here, Minkowski geometry) and we accept that the speed of light
changes from point to point. The Riemann geometry gives us another option. Geometry becomes
the result of observation and the measurement tool may change from point to point. In this case
we can keep the speed of light constant. Geometry becomes a background which depends on
physical processes. Physical laws become background independent.
\fi

\ifx\texPrefaceRefernceFrame\Defined
There are few papers dedicated to a variable speed of light theory
\citeBib{Magueijo,Bassett}. Their theory is based on idea that metric tensor
may be scaleable relative dilatation. This idea is not new.
As soon as Einstein published general relativity
Weyl introduced his idea to make theory invariant relative
conformal transformation. However Einstein was firmly opposed to
this idea because it broke dependence between distance
and proper time. You can find detailed analysis in \citeBib{Straumann}.

We have strong relation between the speed of light
and units of length and time in special and general relativity.
When we develop new theory and discover that speed of light is different
we should ask ourselves about the reason. Did we make accurate measurement?
Do we have alternative way to exchange information and synchronize reference
frame? Did transformations between reference frames change and do they create group?

In some models photon may have small rest mass \citeBib{Lammerzahl}.
In this case speed of light is different from maximal speed and may
depend on direction. Recent experiment \citeBib{Muller}
put limitation on parameters of these models.
\fi

\ifx\texPrefaceTorsion\Defined
\section{Torsion Tensor in general relativity}

Close relationship between the metric tensor and the connection is
the basis of the Riemann geometry.
At the same time, connection and metric as any geometrical object are objects of measurement.
When Hilbert derived Einstein equation, he introduced the lagrangian where
the metric tensor and the connection are independent.
Later on, Hilbert discovered that the connection
is symmetric and fonud dependence between connection and metric tensor. One of the
reasons is in the simplicity of the lagrangian.

\ifx\PrintBook\Defined
Formally chapter \ref{chapter: General Relativistic Field Equations}
is the last chapter in the book. However from historyc aproach
it reflects research which was the starting point of certain period of my life.
Straight join of two lagrangians (of gravitational
and quantum fields) is the heart of equations developed in this chapter.
Equations \eqref{eq: Field}, \eqref{eq: Einstein},
\eqref{eq: Cartan} and
\eqref{eq: Maxwell} have complicated form.
They are nonlinear over connection.
However any attempt to solve these equations lead to case that
this system breaks apart Einstein equation and field equation which are independent.
It is possible that reason of the failure was inside of the method used to solve the problem.
Recent developement of physics prooved that join of
general relativity and quantum mechanics demands more powerful methods.
String theory and loop gravity represent main direction developed today.

However inspite of failure analysis of these equations is very important.
\fi

Since an errors of measurement are inescapable,
analysis of quantum field theory shows that either
symmetry of connection or dependence between connection and metric may be broken. 
This assumption leads to metric\hyph affine manifold
which is space with torsion and nonzero covariant derivative
of metric (section \ref{section: Metric-affine Manifold}).
Independence of the metric tensor and the connection allows us to see
which object is responsible for different
phenomenon in geometry and therefore in physics.
Even we do not prove empirically existence of torsion and nonmetricity
we see here very interesting geometry.

The metric\hyph affine manifold appears in different physical applications.
It is very important to understand what kind of geometry of this space is,
how torsion influences on physical phenomena.
This is why small group of physicists continue to study gravitation
theory with torsion
\citeBib{Mielke,Obukhov,Sardanashvily,Gauge,Neeman}.

In particular we have two different definitions of a geodesic curve in the Riemann manifold.
We consider a geodesic curve either as line of extreme length
(such line is called extreme),
or as line such that tangent vector keep to be tangent to line during parallel transfer
(such line is called auto parallel).
Nonmetricity means that parallel transport does not conserve a length of vector and an angle between
vectors. This leads to a difference between
definitions of auto parallel and extreme lines (\citeBib{torsion} and
section \ref{section: Line with extreme length})
and to a change in the expression of the Frenet transport.
The change of geometry influences the second Newton law which we study
in section \ref{section: Newton Laws: Scalar Potential}.
I show in theorems \ref{theorem: LengthTangentVector} and \ref{theorem: First Newton law} 
that a free falling particle chooses an extreme line transporting
its momentum along the trajectory without change.

Patern of the Newton second law depends on choice of potential.
In case of scalar potential
the Newton second law holds the relationship between force, mass and acceleration.
In case of vector potential
analysis of motion in a gravitational field shows that the field-strength tensor
depends on the derivative of the metric tensor.

Nonmetricity dramatically changes law how orthogonal basis moves in space time.
However learning of parallel transport in space with nonmetricity allows us to
introduce the Cartan transport and an introduction of the connection compatible with the metric tensor
(section \ref{section: Cartan Transport}). The Cartan transport
holds the basis orthonormal and this makes it valuable tool
in dynamics (section \ref{section: Newton Laws: Scalar Potential})
because the observer uses an orthonormal basis as his measurement tool.
The dynamics of a particle follows to the Cartan transport.
The question arises from this conclusion.

We can change the connection as we show in section \ref{section: Cartan Transport}.
Why we need to learn manifolds with an arbitrary connection and the metric tensor?
The learning of the metric\hyph affine manifold shows why everything
works well in the Riemann manifold and what changes in a general case.
What kind of different physical phenomena
may result in different connections?
Physical constraints that appear in a model may lead to appearance of a nonmetricity
\citeBib{Gauge,Megged,gr-qc-9604027}.
Because the Cartan transport is the natural mechanism to conserve orthogonality
we expect that we will interpret a deviation of the test particle from the extreme line as
a result of an external to this particle
force\footnote{For instance if we extend the definition \eqref{eq: potential force} of a force
to a general case \eqref{eq: Newton} we can interpret a deviation of a charged particle
in electromagnetic field as result of the force
\[\ F^j=\frac e {cu^0} g^{ij} F_{li} u^l\]
The same way we can interpret a deviation of the auto parallel line as the force
\[F^i=-\frac {m c} {u^0} \Gamma(C)^i_{kl}u^ku^l\]
I remind that the Cartan symbol is the tensor}.
In this case the difference between two types of a transport
becomes measurable and meaningful. Otherwise another type
of a transport and nonmetricity are not observable and we can use
only the transport compatible with metric.

I see here one more opportunity. As follows from the paper \citeBib{Megged}
the torsion may depend on quantum properties of matter.
However the torsion is the part of the connection. This means that the connection may
also depend on quantum properties of matter. This may lead to breaking of
the Cartan transport. However this opportunity demands additional research.

The effects of torsion and a nonmetricity are cumulative.
They may be small but measurable. We can observe their effects not only in strong
fields like black hole or Big Bang but in regular conditions
as well.
Studying geometry and dynamics of point particle gives us a way to test this point of view.
There is mind to test this theory in condition when spin of quantum field is accumulated.
We can test a deviation from second Newton law or measure torsion by observing
the movement of two different particles.

To test if the spacetime has the torsion we
can test the opportunity to build a parallelogram in spacetime. 
We can get two particles or two photons that start their movement from the same point
and using a mirror to force them to move along opposite sides of the parallelogram.
We can start this test when we do not have quantum field and then repeat
the test in the presence of quantum field.
If particles meet in the same place or we have the same interference then we have
torsion equal $0$ in this thread.
In particular, the torsion may influence the behavior of virtual particles.
\fi

\ifx\PrintBook\Defined
\section{Symmetry}

The torsion leads to a change in the Killing equation. We also need to add a similar equation
for the connection.

Asymmetry of connection arises not only from torsion. Theory of
geometrical object in vector space has natural
generalization: the theory of reference frame in event space. Reference frame
is powerful tool of measurement in general relativity.
\fi

\ifx\texPrefaceTidal\Defined

\section{Tidal Acceleration}

Observations in Solar system and outside are very important. They give us an opportunity
to see where general relativity is right and to find out its limitation.
It is very important to be very careful with such observations.
NASA provided very interesting observation of Pioneer 10 and Pioneer 11
and managed complicated calculation of their accelerations. However, one interesting
question arises:
what kind of acceleration did we measure?

Pioneer 10 and Pioneer 11 performed free movement in solar system.
Therefore they move along their trajectory without acceleration.
However, it is well known that two bodies moving along close geodesics have relative
acceleration that we call tidal acceleration. Tidal acceleration in general relativity has form
	\begin{equation}
	\label{eq: Tidal acceleration}
\frac{D^2\delta x^k}{ds^2}
=R^k_{lnk}\delta x^kv^nv^l
	\end{equation}
where $v^l$ is the speed of body $1$ and $\delta x^k$ is the deviation of geodesic of body $2$
from geodesic of body $1$. We see from this expression that tidal acceleration
depends on movement of body $1$ and how the trajectory of body $2$ deviates from the trajectory of body $1$.
But this means that even for two bodies that are at the same distance from a central body we can measure
different acceleration relative an observer.

Section \ref{section: Tidal Equation} is dedicated to the problem
what kind of changes the tidal force experiences on
metric affine manifolds.

Finally the question arises. Can we use equation \eqref{eq: deviation_extreme}
to measure torsion? We get tidal acceleration from direct measurement. There is
a method to measure curvature (see for instance \citeBib{Wheeler}). However, even
if we know the acceleration and curvature we still have differential equation
to find torsion. However, this way may give direct answer to the question
of whether torsion exists or not.

Deviation from tidal acceleration \eqref{eq: Tidal acceleration}
predicted by general relativity may have different reason.
However we can find answer by combining different type of measurement.
\fi

\def\texGeomObject{}
\ifx\PrintBook\Defined
				\chapter{Geometrical Object}
\fi

			\section{Metric-affine Manifold}
			\label{section: Metric-affine Manifold}

For connection \xEqRef{0405.027}{eq: GLn connection form}
we defined the \AddIndex{torsion form}{torsion form}
	\begin{equation}
T^a = d^2 x^a + \omega^a_b \wedge dx^b
	\label{eq: Torsion}
	\end{equation}
From \xEqRef{0405.027}{eq: GLn connection form} it follows
	\begin{equation}
\omega^a_b \wedge dx^b=(\Gamma^a_{bc}-\Gamma^a_{cb})dx^c\wedge dx^b
	\label{eq: wedge connection}
	\end{equation}
Putting \eqref{eq: wedge connection} and \xEqRef{0405.027}{eq: anholonomity} into \eqref{eq: Torsion}
we get
	\begin{equation}
T^a=T^a_{cb}dx^c\wedge dx^b =
-c^a_{cb}dx^c\wedge dx^b + (\Gamma^a_{bc}-\Gamma^a_{cb})dx^c\wedge dx^b
\label{eq: Torsion 1}
	\end{equation}
where we defined \AddIndex{torsion tensor}{torsion tensor}
	\begin{equation}
T^a_{cb} =
\Gamma^a_{bc}-\Gamma^a_{cb}-c^a_{cb}
\label{eq: Torsion coordinates}
	\end{equation}

Commutator of second derivatives has form
	\begin{equation}
u^\alpha_{;kl}-u^\alpha_{;lk}
=R^\alpha_{\beta lk}u^\beta
-T^p_{lk}u^\alpha_{;p}
	\label{eq: commutator second derivative}
	\end{equation}
From \eqref{eq: commutator second derivative} it follows that
	\begin{equation}
	\label{eq: commutator second derivative of vector}
\xi^a_{;cb}-\xi^a_{;bc}
=R^a_{d bc}\xi^d
-T^p_{bc}\xi^a_{;p}
	\end{equation}

In Rieman space we have metric tensor $g_{ij}$ and connection $\Gamma^k_{ij}$.
One of the features of the Rieman space is symetricity of connection and covariant derivative
of metric is 0. This creates close relation between metric and connection.
However the connection is not necessarily symmetric
and the covariant derivative of the metric tensor may be different from 0.
In latter case we introduce the \AddIndex{nonmetricity}{nonmetricity}
	\begin{equation}
Q^{ij}_k=g^{ij}_{;k}=g^{ij}_{,k}+\Gamma^i_{pk}g^{pj}+\Gamma^j_{pk}g^{ip}
	\label{eq: Nonmetricity}
	\end{equation}

Due to the fact that derivative of the metric tensor is not 0, we cannot raise or lower index
of a tensor under derivative as we do it in regular Riemann space.
Now this operation changes to next
\[a^i_{;k} = g^{ij} a_{j;k} + g^{ij}_{;k} a_j\]
This equation for the metric tensor gets the following form
\[g^{ab}_{;k} = - g^{ai} g^{bj} g_{ij;k}\]

		\begin{definition}
We call a manifold with a torsion
and a nonmetricity
the \AddIndex{metric\Hyph affine manifold}{metric-affine manifold} \citeBib{Mielke}.
		\qed  
		\end{definition}

If we study a submanifold $V_n$ of a manifold $V_{n+m}$, we see that the immersion creates
the connection $\Gamma^\alpha_{\beta\gamma}$ that relates to the connection in manifold as
\[
\Gamma^\alpha_{\beta\gamma} e^l_\alpha =
\Gamma^l_{mk} e^m_\beta e^k_\gamma + \frac {\partial e^l_\beta} {\partial u^\gamma}
\]
Therefore there is no smooth immersion
of a space with torsion into the Riemann space.

			\section{Geometrical Meaning of Torsion}
Suppose that $a$ and $b$ are non collinear vectors in a point $A$ (see figure \ref{fig:Torsion}).

\begin{tabular}{l|r}
\begin{minipage}{160pt}
We draw the geodesic $L_a$ through the point $A$
using the vector $a$ as a tangent vector to $L_a$ in the point $A$.
Let $\tau$ be the canonical parameter on $L_a$ and
\[\frac {dx^k}{d\tau}=a^k\]
We transfer the vector $b$ along the geodesic $L_a$ from the point $A$
into a point $B$ that defined by any value of the parameter $\tau=\rho>0$. We mark the result as $b'$.
\end{minipage}
&
\begin{minipage}{160pt}
We draw the geodesic $L_b$ through the point $A$
using the vector $b$ as a tangent vector to $L_b$ in the point $A$.
Let $\varphi$ be the canonical parameter on $L_b$ and
\[\frac {dx^k}{d\varphi}=b^k\]
We transfer the vector $a$ along the geodesic $L_b$ from the point $A$
into a point $D$ that defined by any value of the parameter $\varphi=\rho>0$. We mark the result as $a'$.
\end{minipage}
\end{tabular}

\begin{tabular}{l|r}
\begin{minipage}{160pt}
We draw the geodesic $L_{b'}$ through the point $B$
using the vector $b'$ as a tangent vector to $L_{b'}$ in the point $B$.
Let $\varphi'$ be the canonical parameter on $L_{b'}$ and 
\[\frac {dx^k}{d\varphi'}=b'^k\]
We define a point $C$ on the geodesic $L_{b'}$ by parameter value $\varphi'=\rho$
\end{minipage}
&
\begin{minipage}{160pt}
We draw the geodesic $L_{a'}$ through the point $D$
using the vector $a'$ as a tangent vector to $L_{a'}$ in the point $D$.
Let $\tau'$ be the canonical parameter on $L_{a'}$ and 
\[\frac {dx^k}{d\tau'}=a'^k\]
We define a point $E$ on the geodesic $L_{a'}$ by parameter value $\tau'=\rho$
\end{minipage}
\end{tabular}

Formally lines $AB$ and $DE$ as well as lines $AD$ and $BC$ are parallel
lines. Lengths of $AB$ and $DE$ are the same as well as lengths
of $AD$ and $BC$ are the same. We call this figure
a \AddIndex{parallelogram}{parallelogram} based on vectors
$a$ and $b$ with the origin in the point $A$.

		\begin{theorem}
Suppose $CBADE$ is a parallelogram with a origin in
the point $A$; then the resulting figure will not be closed \citeBib{torsion}.
The value of the difference of coordinates
of points $C$ and $E$ is equal to surface integral of the torsion
over this parallelogram\footnote{Proof of this statement I found in \citeBib{Shilov}}
\[\Delta_{CE}x^k=\iint T^k_{mn}dx^m \wedge dx^n \]
		\end{theorem}

	\begin{figure}
	\begin{center}
	\begin{picture}(150,150)
\put( 20, 25){ $A$ }
\put( 30, 65){ $a$ }
\put( 45, 50){ $b$ }
\put( 55, 130){ $B$ }
\put( 80, 135){ $b'$ }
\put( 155, 80){ $a'$ }
\put( 165, 150){ $C$ }
\put( 170, 40){ $D$ }
\put( 195, 135){ $E$ }
\put(-115, -65){\qbezier(150,100)(162,150)(185,190)}
\put(35, 35){\vector(1,4){11}}
\put(-115, -65){\qbezier(185,190)(218,207)(283,210)}
\put(70, 125){\vector(2,1){30}}
\put(169, 145){\line(3, -2){18}}
\put(-149, -154){\qbezier(320,212)(323,260)(336,287)}
\put(171, 59){\vector(0,1){35}}
\put(-149, -154){\qbezier(185,190)(208,206)(320,212)}
\put(35, 35){\vector(3,2){30}}
	\end{picture}
	\caption{Meaning of Torsion}
	\label{fig:Torsion}
	\end{center}
	\end{figure}
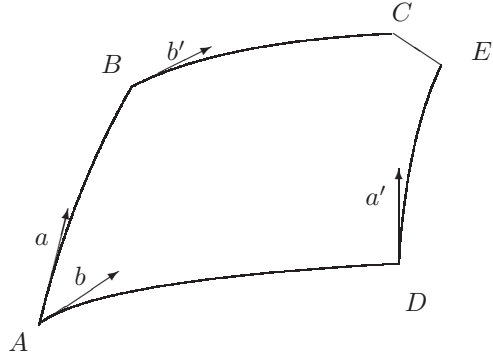
		\begin{proof}
We can find an increase of coordinate $x^k$ along any geodesic as
	\begin{align*}
\Delta x^k&=\frac{dx^k}{d\tau}\tau+\frac 1 2 \frac{d^2 x^k}{d\tau^2}\tau^2+O(\tau^2)=\\
&=\frac{dx^k}{d\tau}\tau-\frac 1 2 \Gamma^k_{mn}\frac{dx^m}{d\tau} \frac{dx^n}{d\tau}\tau^2+O(\tau^2)
	\end{align*}
where $\tau$ is canonical parameter and we take values of derivatives and components
$\Gamma^k_{mn}$ in the initial point. In particular we have 
	\[
\Delta_{AB} x^k=a^k\rho-\frac 1 2 \Gamma^k_{mn}(A)a^m a^n\rho^2+O(\rho^2)
	\]
along the geodesic $L_a$ and
	\begin{equation}
\Delta_{BC} x^k=b'^k\rho-\frac 1 2 \Gamma^k_{mn}(B)b'^m b'^n\rho^2+O(\rho^2)
	\label{eq:Torsion_1}
	\end{equation}
along the geodesic $L_{b'}$. Here
	\begin{equation}
b'^k=b^k- \Gamma^k_{mn}(A)b^m dx^n+O(dx)
	\label{eq:Torsion_2}
	\end{equation}
is the result of parallel transfer of $b^k$ from $A$ to $B$ and
	\begin{equation}
dx^k=\Delta_{AB} x^k=a^k\rho
	\label{eq:Torsion_3}
	\end{equation}
with precision of small value of first level. Putting
\eqref{eq:Torsion_3} into \eqref{eq:Torsion_2} and \eqref{eq:Torsion_2} into
\eqref{eq:Torsion_1} we will receive
\[
\Delta_{BC} x^k=b^k \rho- \Gamma^k_{mn}(A)b^m a^n \rho^2
-\frac 1 2 \Gamma^k_{mn}(B)b^m b^n\rho^2+O(\rho^2)
\]
Common increase of coordinate $x^K$ along the way $ABC$ has form
\[
\Delta_{ABC}x^k=\Delta_{AB} x^k+\Delta_{BC} x^k=
\]
	\begin{equation}
	\label{eq:Torsion_4}
=(a^k+b^k)\rho- \Gamma^k_{mn}(A)b^m a^n \rho^2-
	\end{equation}
\[
-\frac 1 2 \Gamma^k_{mn}(B)b^m b^n\rho^2
-\frac 1 2 \Gamma^k_{mn}(A)a^m a^n\rho^2+O(\rho^2)
\]

Similar way common increase of coordinate $x^K$ along the way $ADE$ has form
\[
\Delta_{ADE}x^k=\Delta_{AD} x^k+\Delta_{DE} x^k=
\]
	\begin{equation}
	\label{eq:Torsion_5}
=(a^k+b^k)\rho- \Gamma^k_{mn}(A)a^m b^n \rho^2-
	\end{equation}
\[
-\frac 1 2 \Gamma^k_{mn}(D)a^m a^n\rho^2
-\frac 1 2 \Gamma^k_{mn}(A)b^m b^n\rho^2+O(\rho^2)
\]

From \eqref{eq:Torsion_4} and \eqref{eq:Torsion_5}, it follows that
\[
\Delta_{ADE}x^k-\Delta_{ABC}x^k=\]
\[= \Gamma^k_{mn}(A)b^m a^n\rho^2
+\underline{\frac 1 2 \Gamma^k_{mn}(B)b^m b^n\rho^2}_1
+\underline{\frac 1 2 \Gamma^k_{mn}(A)a^m a^n\rho^2}_2-\]
\[- \Gamma^k_{mn}(A)a^m b^n \rho^2
-\underline{\frac 1 2 \Gamma^k_{mn}(D)a^m a^n\rho^2}_2
-\underline{\frac 1 2 \Gamma^k_{mn}(A)b^m b^n\rho^2}_1+O(\rho^2)
\]
For small enough value of $\rho$ underlined terms annihilate each other
and we get integral sum for expression
\[
\Delta_{ADE}x^k-\Delta_{ABC}x^k=\iint_\Sigma (\Gamma^k_{nm}- \Gamma^k_{mn})dx^m \wedge dx^n
\]

However it is not enough to find the difference
\[\Delta_{ADE}x^k-\Delta_{ABC}x^k\]
to find the difference of coordinates of points $C$ and $E$.
Coordinates may be anholonomic and we have to consider that
coordinates along closed loop change \xEqRef{0405.027}{eq: change of coordinate along a loop}
\[\Delta x^k=\oint_{ECBADE}dx^k=-\iint_\Sigma c^k_{mn}dx^m \wedge dx^n \]
where $c$ is anholonomity object.

Finally the difference of coordinates of points $C$ and $E$ is
\[\Delta_{CE}x^k=\Delta_{ADE}x^k-\Delta_{ABC}x^k+\Delta x^k=
\iint_\Sigma (\Gamma^k_{nm}- \Gamma^k_{mn}-c^k_{mn})dx^m \wedge dx^n\]
Using \eqref{eq: Torsion coordinates} we prove the statement. 
		\end{proof}

				\section{Relation between Connection and Metric}

Now we want to find how we can express connection if we know
metric and torsion. According to definition
\[-Q_{kij}=g_{ij;k}=g_{ij,k}-\Gamma^p_{ik}g_{pj}-\Gamma^p_{jk}g_{pi}\]
\[-Q_{kij}=g_{ij,k}-\Gamma^p_{ik}g_{pj}-\Gamma^p_{kj}g_{pi}-S^p_{jk}g_{pi}\]
We move derivative of $g$ and torsion to the left-hand side.
	\begin{equation}
g_{ij,k}+Q_{kij}-S^p_{jk}g_{pi}=\Gamma^p_{ik}g_{pj}+\Gamma^p_{kj}g_{pi}
	\label{eq: Connection1}
	\end{equation}
Changing order of indexes we write two more equations
	\begin{equation}
g_{jk,i}+Q_{ijk}-S^p_{ki}g_{pj}=\Gamma^p_{ji}g_{pk}+\Gamma^p_{ik}g_{pj}
	\label{eq: Connection2}
	\end{equation}
	\begin{equation}
g_{ki,j}+Q_{jki}-S^p_{ij}g_{pk}=\Gamma^p_{kj}g_{pi}+\Gamma^p_{ji}g_{pk}
	\label{eq: Connection3}
	\end{equation}
If we substruct equation \eqref{eq: Connection1} from sum of equations
\eqref{eq: Connection2} and \eqref{eq: Connection1} we get
\[
g_{ki,j}+g_{jk,i}-g_{ij,k}+Q_{ijk}+Q_{jki}-Q_{kij}
-S^p_{ij}g_{pk}-S^p_{ki}g_{pj}+S^p_{jk}g_{pi}=2\Gamma^p_{ji}g_{pk}
\]
Finally we get
\[
\Gamma^p_{ji}=\frac 1 2 g^{pk}
(g_{ki,j}+g_{jk,i}-g_{ij,k}+Q_{ijk}+Q_{jki}-Q_{kij}
-S^r_{ij}g_{rk}-S^r_{ki}g_{rj}+S^r_{jk}g_{ri})
\]

\def\texAffine{}
\ifx\PrintBook\undefined
\else
				\chapter{Geometry of Metric-Affine Manifold}
\fi

			\section{Line with Extreme Length}
			\label{section: Line with extreme length}

There are two different definitions of a geodesic curve in the Riemann manifold.
One of them relies on the parallel transport. We call an appropriate \AddIndex{line auto parallel}{auto parallel line}.
Another definition depends on the length of trajectory. We call an appropriate \AddIndex{line extreme}{extreme line}.
In a metric\hyph affine manifold these lines have different equations \citeBib{torsion}.
Equation of auto parallel line does not change. However, the equation of extreme
line changes\footnote{To derive the equation \eqref{eq: xtremeLine}
I follow the ideas that Rashevsky \citeBib{Rashevsky}
implemented for the Riemann manifold}.

		\begin{theorem}
Let $x^i = x^i(t,\alpha)$ be a line depending on a parameter $\alpha$ with fixed points at $t = t_1$ and
$t = t_2$ and we define its length as
\begin{equation}
s = \int^{t_2}_{t_1} \sqrt{g_{ij} \frac {dx^i} {dt} \frac {dx^j} {dt}} dt
\label{eq: LineLength}
\end{equation}
Then 
\begin{equation}
\delta s =
\int^{t_2}_{t_1} \left( \frac 1 2 \left(
g_{kj;i}- g_{ik;j}
- g_{ij;k} \right) \frac{dx^k}{ds} \frac{dx^j}{ds} ds -
g_{ij} D\frac {dx^j} {ds} \right) \delta x^i
\label{eq: VarLineLength}
\end{equation}
where $\delta x^k$ is the change of a line when $\alpha$ changes.
		\end{theorem}
		\begin{proof}
We have
$$\frac {ds} {dt} = \sqrt{g_{ij} \frac {dx^i} {dt} \frac {dx^j} {dt}}$$
and
\[
\delta s = \int^{t_2}_{t_1}
\frac
 {\delta \left( g_{ij} \frac{dx^i}{dt} \frac{dx^j}{dt} \right)}
 {2 \frac{ds}{dt}} dt
\]

We can estimate the numerator of this fraction as
\[
g_{ij,k} \delta x^k \frac {dx^i} {dt} \frac {dx^j} {dt} +
2 g_{ij} \delta \frac {dx^i} {dt} \frac {dx^j} {dt} =
\]
\[
= g_{ij;k} \delta x^k {dx^i} {dt} \frac {dx^j} {dt} +
2 g_{ij} \Gamma^i_{lk} \delta x^k \frac {dx^l} {dt} \frac {dx^j} {dt}
+2 g_{ij} d \frac {\delta x^i} {dt} \frac {dx^j} {dt} =
\]
\[
= g_{ij;k} \delta x^k \frac{dx^i}{dt} \frac{dx^j}{dt} + 2 g_{ij} \frac{D\delta x^i}{dt} \frac{dx^j}{dt}
\]
and we have
\[
\delta s = \int^{t_2}_{t_1}
{
 \frac
 {
  g_{ij;k} \delta x^k dx^i \frac{dx^j}{dt} + 2 g_{ij} D\delta x^i \frac{dx^j}{dt}
 } 
 {2 \frac {ds} {dt}}
}
\]
\[
= \int^{t_2}_{t_1}
\left(\frac 1 2 g_{ij;k} \delta x^k dx^i \frac {dx^j} {ds} +
g_{ij} D\delta x^i \frac {dx^j} {ds} \right) 
\]
\[
= \int^{t_2}_{t_1}
\left(\frac 1 2 g_{kj;i} \delta x^i \frac {dx^k} {ds} {ds} \frac {dx^j} {ds} +
d\left(g_{ij} \delta x^i \frac {dx^j} {ds}\right)
-g_{ij;k} \frac {dx^k} {ds} {ds} \frac {dx^j} {ds} \delta x^i -
g_{ij} D\frac {dx^j} {ds} \delta x^i\right) \]
\[
= 
\left.\left(g_{ij} \delta x^i \frac{dx^j}{ds} \right)\right|^{t_2}_{t_1}
+ \int^{t_2}_{t_1} \left(\frac 1 2 \left(
g_{kj;i} -
g_{ij;k} -
g_{ik;j} \right) \frac {dx^k} {ds} \frac{dx^j}{ds} {ds}
- g_{ij} D\frac{dx^j}{ds}\right) \delta x^i
\]
First term is $0$ because points, when $t = t_1$ and $t = t_2$, are fixed.
Therefore, we have got the statement of the theorem.
		\end{proof}

		\begin{theorem}
		\label{theorem: xtremeLine}
An extreme line satisfies equation
	\begin{equation}
\label{eq: xtremeLine}
\frac {D\frac {dx^l} {ds}} {ds} =
\frac 1 2 g^{il} \left(g_{kj;i} - g_{ik;j}
- g_{ij;k}\right) \frac{dx^k}{ds} \frac{dx^j}{ds}
	\end{equation}
		\end{theorem}
		\begin{proof}
To find a line with extreme length we use the functional \eqref{eq: LineLength}.
Since $\delta s = 0$,
$$
\frac 1 2 \left(g_{kj;i} -
g_{ij;k} -
g_{ik;j}\right) \frac {dx^k} {ds} \frac {dx^j} {ds} ds
- g_{ij} D\frac {dx^j} {ds} = 0$$
follows from \eqref{eq: VarLineLength}.
		\end{proof}

		\begin{theorem}
Parallel transport along an extreme line holds length of tangent vector.
		\end{theorem}

		\begin{proof}

Let $$v^i=\frac {dx^i}{ds}$$ be the tangent vector to extreme curve.
From theorem \ref{theorem: xtremeLine} it follows that
$$\frac {Dv^l} {ds} =
g^{il} \frac 1 2 \left(g_{kj;i} - g_{ik;j}
- g_{ij;k}\right) v^k v^j
$$
and
$$\frac {Dg_{kl} v^k v^l}{ds}
= \frac {Dg_{kl}} {ds}v^k v^l
+ g_{kl}\frac {Dv^k} {ds} v^l
+ g_{kl}v^k \frac {Dv^l} {ds}=$$
$$= g_{kl;p}v^p v^k v^l+$$
$$+ g_{kl}g^{ik} \frac 1 2 \left(g_{rj;i} - g_{ir;j}
- g_{ij;r}\right) v^r v^j v^l
+ g_{kl}v^k g^{il} \frac 1 2 \left(g_{rj;i} - g_{ir;j}
- g_{ij;r}\right) v^r v^j=$$
$$= g_{kl;p}v^p v^k v^l
+ \left(g_{rj;l} - g_{lr;j}
- g_{lj;r}\right) v^r v^j v^l = 0$$
Therefore length of the vector $v^i$ does not change along extreme curve.
		\end{proof}

			\section{Frenet Transport}
			\label{section: Frenet Transport}

All equations that we derived before are different,
however they have something common in their structure.
All these equations express movement along a line and in the right side of them
we can see the curvature of this line.

By definition curvature of a line is
\[\xi(s) = \left| \frac {D\frac {dx^l} {ds}} {ds} \right|\]
Therefore we can introduce unit vector $e_1$ such that
\[\frac {D\frac {dx^l} {ds}} {ds} = \xi e^l_1\]

Knowledge of the transport of a basis along a line is very important,
because it allows us to study how spacetime changes when an observer moves through it.
Our task is to discover equations similar to the Frenet transport in
the Riemann space. We design the accompaniment basis $\nu^i_k$ the same
way we do it in the Riemann space.

Vectors
\[\xi^i(t)=\frac{dx^i(t)}{dt}, \ \ \ \frac {D\xi^i} {dt},
\ \ \ ...\ \ \ \frac {D^{n-1}\xi^i} {dt^{n-1}}\]
in general are linearly independent. We call plane that we
create on the base of first $p$ vectors as $p$-th osculating plane $R_p$.
This plane does not depend on choice of parametr $t$.

Our next task is to create orthogonal basis which shows us how line changes.
We get vector $\nu^i_1\in R_1$ so it is tangent to line.
We get vector $\nu^i_p\in R_p$, $p>1$ such that $\nu^i_p$ is
orthogonal to $R_{p-1}$. If original line is not isotropic
then each $\nu^i_p$ also is not isotropic and we can get unit vector
in the same direction. We call this basis accompaniment.

		\begin{theorem}
		\label{theorem: FrenetTransfer}
The \AddIndex{Frenet transport}{Frenet transport} in the metric\hyph affine manifold gets the form
	\begin{equation}
	\begin{split}
\frac {D\nu^i_p} {dt}
= \frac 1 2 g^{im}(g_{kl;m}-g_{km;l}-g_{ml;k}) \nu^k_1 \nu^l_p - \\
- \epsilon_p \epsilon_{p-1} \xi_{p-1} \nu^i_{p-1} + \xi_p \nu^i_{p+1}
	\end{split}
	\label{eq: Frenet}
	\end{equation}
\[\epsilon_k = sign(g_{pq} \nu^p_k \nu^q_k)\]
Here $\nu^a_k$ is vector of basis, moving along line,
\[\epsilon_k = sign(g_{pq} \nu^p_k \nu^q_k)\]
		\end{theorem}
		\begin{proof}
We introduce vectors $\nu^a_k$ in this way that
	\begin{equation}
\frac {D\nu^i_p} {dt}
= \frac 1 2 g^{im}(g_{kl;m}-g_{km;l}-g_{ml;k}) \nu^k_1 \nu^l_p
+ a^q_p \nu^i_q
	\label{eq: Frenet_1}
	\end{equation}
where $a_p^q = 0$ when $q > p+1$.
Now we can determine coefficients $a^q_p$. If we get derivative of the equation
\[
g_{ij} \nu^i_p \nu^j_q = const
\]
and substitute \eqref{eq: Frenet_1} we get the equation
\[\frac {dg_{ij}\nu^i_a \nu^j_b} {ds}=
\frac {Dg_{ij}} {ds}\nu^i_a \nu^j_b+
g_{ij}\frac {D\nu^i_a} {ds}\nu^j_b+
g_{ij}\nu^i_a\frac {D\nu^j_b} {ds}=\]
\[=g_{ij;k}\nu^k_1 \nu^i_a \nu^j_b+\]
\[+g_{ij}(\frac 1 2 g^{im}(g_{kl;m}-g_{km;l}-g_{ml;k}) \nu^k_1 \nu^l_a
+ a^q_a \nu^i_q)\nu^j_b+\]
\[+g_{ij}\nu^i_a (\frac 1 2 g^{jm}(g_{kl;m}-g_{km;l}-g_{ml;k}) \nu^k_1 \nu^l_b
+ a^q_b \nu^i_q)=\]
\[=g_{ij;k}\nu^k_1 \nu^i_a \nu^j_b+\]
\[+g_{ij}\frac 1 2 g^{im}(g_{kl;m}-g_{km;l}-g_{ml;k}) \nu^k_1 \nu^l_a\nu^j_b
+ g_{ij}a^q_a \nu^i_q\nu^j_b+\]
\[+g_{ij}\nu^i_a \frac 1 2 g^{jm}(g_{kl;m}-g_{km;l}-g_{ml;k}) \nu^k_1 \nu^l_b
+ g_{ij}\nu^i_a a^q_b \nu^i_q=\]
\[=\frac 1 2 \nu^k_1 \nu^i_a \nu^j_b(2g_{ij;k}+
g_{ki;j}-g_{kj;i}-g_{ji;k}+
g_{kj;i}-g_{ki;j}-g_{ij;k})+\]
\[+ \epsilon_b a^b_a+
+ \epsilon_a a^a_b=0\]

$a_p^q = 0$ when $q > p+1$ by definition.
Therefore $a_p^q = 0$ when $q < p-1$.
Introducing $\xi_p = a^{p+1}_p$ we get
\[
a^p_{p+1} = - \epsilon_p \epsilon_{p+1} \xi_p 
\]
When $q = p$ we get
\[
a^p_p = 0
\]
We get \eqref{eq: Frenet} when substitute $a^q_p$ in \eqref{eq: Frenet_1}.
		\end{proof}

			\section{Lie Derivative}
			\label{section: Lie Derivative}

Vector field $\xi^k$ on manifold generates infinitesimal transformation
	\begin{equation}
	\label{eq: infinitesimal displacement}
x'^k=x^k+\epsilon\xi^k
	\end{equation}
which leads to the \AddIndex{Lie derivative}{Lie derivative}.
Lie derivative tells us how the object changes when we move along the vector field.

		\begin{theorem}
\AddIndex{Lie derivative of metric}{Lie derivative of metric} has form
\symb{\mathcal{L}_\xi g_{ab}}0{Lie derivative of metric}
	\begin{equation}
	\label{eq: Lie derivative of metric}
\ShowSymbol{Lie derivative of metric}=\xi^k_{;<a>}g_{kb}+\xi^k_{;<b>}g_{ka}
+T^l_{ka}g_{lb}\xi^k+T^l_{kb}g_{la}\xi^k+g_{ab;<k>}\xi^k
	\end{equation}
		\end{theorem}
		\begin{proof}
We start from transformation \eqref{eq: infinitesimal displacement}.
Then
\[g_{ab}(x')=g_{ab}(x)+g_{ab,c}\epsilon \xi^c\]
	\begin{align*}
g'_{ab}(x')&=\frac{\partial x^c}{\partial x'^a} \frac{\partial x^d}{\partial x'^b}g_{cd}(x)\\
&=g_{ab}-\epsilon \xi^c_{,a}g_{cb}-\epsilon \xi^c_{,b}g_{ac}
	\end{align*}
According to definition of Lie derivative we have
	\begin{align*}
\mathcal{L}_\xi g_{ab}&=g_{ab}(x')-g'_{ab}(x')\\
&=g_{ab,c}\epsilon \xi^c+\epsilon \xi^c_{,a}g_{cb}+\epsilon \xi^c_{,b}g_{ac}\\
&=(g_{ab;<c>}+\overline{\Gamma^d_{ac}}g_{db}+\overline{\Gamma^d_{bc}}g_{ad})\epsilon \xi^c\\
&+\epsilon (\xi^c_{;<a>}-\overline{\Gamma^c_{da}}\xi^d)g_{cb}
+\epsilon (\xi^c_{;<b>}-\overline{\Gamma^c_{db}}\xi^d)g_{ac}
	\end{align*}
	\begin{equation}
	\begin{split}
\mathcal{L}_\xi g_{ab}&=g_{ab;<c>} \xi^c+\overline{\Gamma^d_{ac}}g_{db} \xi^c+\overline{\Gamma^d_{bc}}g_{ad}\xi^c\\
&+ \xi^c_{;<a>}g_{cb}- \overline{\Gamma^c_{da}}\xi^dg_{cb}
+ \xi^c_{;<b>}g_{ac}- \overline{\Gamma^c_{db}}\xi^dg_{ac}
	\end{split}
	\label{eq: Lie derivative of metric,1}
	\end{equation}
\eqref{eq: Lie derivative of metric} follows from
\eqref{eq: Lie derivative of metric,1} and \eqref{eq: Torsion coordinates}.
		\end{proof}
		\begin{theorem}
\AddIndex{Lie derivative of connection}{Lie derivative of connection} has form
\symb{\mathcal{L}_\xi\Gamma^a_{bc}}0{Lie derivative of connection}
	\begin{equation}
	\label{eq: Lie derivative of connection}
\ShowSymbol{Lie derivative of connection}=-\overline{R^a_{bcp}}\xi^p
-T^a_{bp;<c>}\xi^p
-T^a_{be}\xi^e_{;<c>}
+\xi^a_{;<bc>}
	\end{equation}
		\end{theorem}
		\begin{proof}
We start from transformation \eqref{eq: infinitesimal displacement}.
Then
	\begin{equation}
	\begin{split}
\overline{\Gamma^a_{bc}(x')}
&=\Gamma^a_{bc}(x')+A^a_{bc}(x')\\
&=\Gamma^a_{bc}(x)+\Gamma^a_{bc,p}\epsilon \xi^p+A^a_{bc}(x)+A^a_{bc,p}\epsilon \xi^p\\
&=\overline{\Gamma^a_{bc}(x)}+\overline{\Gamma^a_{bc,p}}\epsilon \xi^p
	\end{split}
	\label{eq: Lie derivative of connection,1}
	\end{equation}
	\begin{align*}
\overline{\Gamma'^a_{bc}}(x')
&=\Gamma'^a_{bc}(x')+A'^a_{bc}(x')\\
&=\frac{\partial x'^a}{\partial x^e}\frac{\partial x^f}{\partial x'^b} \frac{\partial x^g}{\partial x'^c}
\Gamma^e_{fg}(x)
+\frac{\partial x'^a}{\partial x^e}\frac{\partial^2 x^e}{\partial x'^b\partial x'^c}
+\frac{\partial x'^a}{\partial x^e}\frac{\partial x^f}{\partial x'^b} \frac{\partial x^g}{\partial x'^c}
A^e_{fg}(x)\\
&=\Gamma^a_{bc}+\epsilon \xi^a_{,e}\Gamma^e_{bc}
-\epsilon \xi^e_{,b}\Gamma^a_{ec}-\epsilon \xi^e_{,c}\Gamma^a_{be}
+(\delta^a_e+\epsilon \xi^a_{,e})(-\epsilon \xi^e_{,cb}))\\
&+A^a_{bc}+\epsilon \xi^a_{,e}A^e_{bc}
-\epsilon \xi^e_{,b}A^a_{ec}-\epsilon \xi^e_{,c}A^a_{be}
	\end{align*}
	\begin{equation}
\overline{\Gamma'^a_{bc}}(x')
=\overline{\Gamma^a_{bc}}+\epsilon \xi^a_{,e}\overline{\Gamma^e_{bc}}
-\epsilon \xi^e_{,b}\overline{\Gamma^a_{ec}}-\epsilon \xi^e_{,c}\overline{\Gamma^a_{be}}
-\epsilon \xi^a_{,cb}
	\label{eq: Lie derivative of connection,2}
	\end{equation}
By definition
\[
\xi^a_{;<e>}
=\xi^a_{,e}
+\overline{\Gamma^a_{pe}} \xi^p
\]
	\begin{equation}
\xi^a_{,e}
=\xi^a_{;<e>}
-\overline{\Gamma^a_{pe}} \xi^p
	\label{eq: Lie derivative of connection,3}
	\end{equation}
	\begin{align*}
\xi^a_{;<ef>}
&=\xi^a_{;<e>,f}
+\overline{\Gamma^a_{pf}} \xi^p_{;<e>}
-\overline{\Gamma^p_{ef}} \xi^a_{;<p>}\\
&=\xi^a_{,ef}+\overline{\Gamma^a_{pe,f}} \xi^p+\overline{\Gamma^a_{pe}} \xi^p_{,f}
+\overline{\Gamma^a_{pf}} \xi^p_{;<e>}
-\overline{\Gamma^p_{ef}} \xi^a_{;<p>}\\
&=\xi^a_{,ef}+\overline{\Gamma^a_{pe,f}} \xi^p
+\overline{\Gamma^a_{pe}} \xi^p_{;<f>}-\overline{\Gamma^a_{pe}} \overline{\Gamma^p_{rf}} \xi^r
+\overline{\Gamma^a_{pf}} \xi^p_{;<e>}
-\overline{\Gamma^p_{ef}} \xi^a_{;<p>}
	\end{align*}
	\begin{equation}
\xi^a_{,ef}
=\xi^a_{;<ef>}-\overline{\Gamma^a_{pe,f}} \xi^p
-\overline{\Gamma^a_{pe}} \xi^p_{;<f>}+\overline{\Gamma^a_{pe}} \overline{\Gamma^p_{rf}} \xi^r
-\overline{\Gamma^a_{pf}} \xi^p_{;<e>}
+\overline{\Gamma^p_{ef}} \xi^a_{;<p>}
	\label{eq: Lie derivative of connection,4}
	\end{equation}
We substitute \eqref{eq: Lie derivative of connection,4} and \eqref{eq: Lie derivative of connection,3}
into \eqref{eq: Lie derivative of connection,2} and get
	\begin{align*}
&\overline{\Gamma'^a_{bc}}(x')\\
&=\overline{\Gamma^a_{bc}}+\epsilon (\underline{\xi^a_{;<e>}}_{4:T}
-\overline{\Gamma^a_{pe}} \xi^p)\overline{\Gamma^e_{bc}}
-\epsilon (\underline{\xi^e_{;<b>}}_2
-\underline{\overline{\Gamma^e_{pb}} \xi^p}_1)\overline{\Gamma^a_{ec}}-\epsilon (\underline{\xi^e_{;<c>}}_{3:T}
-\overline{\Gamma^e_{pc}} \xi^p)\overline{\Gamma^a_{be}}\\
&-\epsilon (\xi^a_{;<cb>}-\overline{\Gamma^a_{pc,b}} \xi^p
-\underline{\overline{\Gamma^a_{pc}} \xi^p_{;<b>}}_2
+\underline{\overline{\Gamma^a_{pc}} \overline{\Gamma^p_{rb}} \xi^r}_1
-\underline{\overline{\Gamma^a_{pb}} \xi^p_{;<c>}}_3
+\underline{\overline{\Gamma^p_{cb}} \xi^a_{;<p>}}_4)
	\end{align*}
	\begin{equation}
\overline{\Gamma'^a_{bc}}(x')
=\overline{\Gamma^a_{bc}}+\epsilon (\xi^a_{;<e>}T^e_{cb}
-\overline{\Gamma^a_{pe}} \xi^p\overline{\Gamma^e_{bc}}
+\xi^e_{;<c>}T^a_{be}
+\overline{\Gamma^e_{pc}} \xi^p\overline{\Gamma^a_{be}}
-\xi^a_{;<cb>}+\overline{\Gamma^a_{pc,b}} \xi^p)
	\label{eq: Lie derivative of connection,5}
	\end{equation}
According definition of Lie derivative we have
using \eqref{eq: Lie derivative of connection,1} and \eqref{eq: Lie derivative of connection,5}
	\begin{align*}
\mathcal{L}_\xi\overline{\Gamma^a_{bc}}
&=(\overline{\Gamma^a_{bc}}(x')-\overline{\Gamma'^a_{bc}}(x'))\epsilon^{-1}\\
&=(\overline{\Gamma^a_{bc}}+\overline{\Gamma^a_{bc,p}}\epsilon \xi^p\\
&-\overline{\Gamma^a_{bc}}-\epsilon (\xi^a_{;<e>}T^e_{cb}
-\overline{\Gamma^a_{pe}} \xi^p\overline{\Gamma^e_{bc}}
+\xi^e_{;<c>}T^a_{be}
+\overline{\Gamma^e_{pc}} \xi^p\overline{\Gamma^a_{be}}
-\xi^a_{;<cb>}+\overline{\Gamma^a_{pc,b}} \xi^p))\epsilon^{-1}
	\end{align*}
	\begin{equation}
\mathcal{L}_\xi\overline{\Gamma^a_{bc}}=\overline{\Gamma^a_{bc,p}} \xi^p
-\xi^a_{;<e>}T^e_{cb}
+\overline{\Gamma^a_{pe}} \xi^p\overline{\Gamma^e_{bc}}
-\xi^e_{;<c>}T^a_{be}
-\overline{\Gamma^e_{pc}} \xi^p\overline{\Gamma^a_{be}}
+\xi^a_{;<cb>}-\overline{\Gamma^a_{pc,b}} \xi^p
	\label{eq: Lie derivative of connection,6}
	\end{equation}
From \eqref{eq: Lie derivative of connection,6}
and \eqref{eq: Torsion coordinates} it follows
	\begin{align*}
\mathcal{L}_\xi\overline{\Gamma^a_{bc}}
&=\overline{\Gamma^a_{cb,p}} \xi^p-\overline{\Gamma^a_{cp,b}} \xi^p\\
&+\underline{\overline{\Gamma^a_{pe}} \overline{\Gamma^e_{bc}}\xi^p}_{3:T}
-\underline{\overline{\Gamma^a_{ep}} \overline{\Gamma^e_{bc}}\xi^p}_3
+\underline{\overline{\Gamma^a_{ep}} \overline{\Gamma^e_{bc}}\xi^p}_{4:T}
-\underline{\overline{\Gamma^a_{ep}} \overline{\Gamma^e_{cb}}\xi^p}_4
+\overline{\Gamma^a_{ep}} \overline{\Gamma^e_{cb}}\xi^p\\
&-\underline{\overline{\Gamma^e_{pc}} \overline{\Gamma^a_{be}}\xi^p}_{1:T}
+\underline{\overline{\Gamma^e_{pc}} \overline{\Gamma^a_{eb}}\xi^p}_1
- \underline{\overline{\Gamma^a_{eb}}\overline{\Gamma^e_{pc}}\xi^p}_{2:T}
+\underline{\overline{\Gamma^a_{eb}}\overline{\Gamma^e_{cp}}\xi^p}_2
-\overline{\Gamma^a_{eb}}\overline{\Gamma^e_{cp}}\xi^p\\
& -\xi^a_{;<e>}T^e_{cb}
-\xi^e_{;<c>}T^a_{be}
+\xi^a_{;<cb>}-T^a_{cp,b} \xi^p-T^a_{bc,p} \xi^p
	\end{align*}
	\begin{equation}
	\begin{split}
\mathcal{L}_\xi\overline{\Gamma^a_{bc}}
&=\overline{\Gamma^a_{cb,p}} \xi^p-\overline{\Gamma^a_{cp,b}} \xi^p
+\overline{\Gamma^a_{ep}} \overline{\Gamma^e_{cb}}\xi^p
-\overline{\Gamma^a_{eb}}\overline{\Gamma^e_{cp}}\xi^p\\
&-\underline{T^a_{pe} \overline{\Gamma^e_{bc}}\xi^p}_{4:T}-\underline{\overline{\Gamma^a_{ep}} T^e_{bc}\xi^p}_1
-\underline{\overline{\Gamma^e_{pc}} T^a_{eb}\xi^p}_{3:T}- \underline{\overline{\Gamma^a_{eb}}T^e_{cp}\xi^p}_2\\
& -\xi^a_{;<e>}T^e_{cb}
-\xi^e_{;<c>}T^a_{be}+\xi^a_{;<cb>}\\
&-T^a_{cp;<b>} \xi^p
+\underline{\overline{\Gamma^a_{eb}} T^e_{cp}\xi^p}_2
-\underline{\overline{\Gamma^e_{cb}} T^a_{ep}\xi^p}_4
-\underline{\overline{\Gamma^e_{pb}} T^a_{ce}\xi^p}_{5:T}\\
&-T^a_{bc;<p>} \xi^p
+\underline{\overline{\Gamma^a_{ep}} T^e_{bc}\xi^p}_1-\underline{\overline{\Gamma^e_{bp}} T^a_{ec}\xi^p}_5
-\underline{\overline{\Gamma^e_{cp}} T^a_{be}\xi^p}_3
	\end{split}
	\label{eq: Lie derivative of connection,7}
	\end{equation}
From \eqref{eq: Lie derivative of connection,7} and \xEqRef{0405.027}{eq: GLn curvature} it follows
	\[
	\begin{split}
\mathcal{L}_\xi\overline{\Gamma^a_{bc}}&=\overline{R^a_{cpb}} \xi^p\\
&-T^e_{cp} T^a_{eb}\xi^p
-T^a_{pe} T^e_{cb}\xi^p-T^e_{bp} T^a_{ce}\xi^p\\
& -\xi^a_{;<e>}T^e_{cb}
-\xi^e_{;<c>}T^a_{be}+\xi^a_{;<cb>}\\
&-T^a_{cp;<b>} \xi^p
-T^a_{bc;<p>} \xi^p
	\end{split}
	\]
	\begin{equation}
	\label{eq: Lie derivative of connection,8}
	\begin{split}
\mathcal{L}_\xi\overline{\Gamma^a_{bc}}&=-\overline{R^a_{cbp}} \xi^p\\
&- (T^a_{eb}T^e_{cp}
+T^a_{ep} T^e_{bc}+ T^a_{ec}T^e_{pb})\xi^p\\
& -\xi^a_{;<e>}T^e_{cb}
-\xi^e_{;<c>}T^a_{be}+\xi^a_{;<cb>}\\
&-T^a_{cp;<b>} \xi^p
-T^a_{bc;<p>} \xi^p
	\end{split}
	\end{equation}
From \eqref{eq: Lie derivative of connection,8} and \eqref{eq: Bianchi 1} it follows
	\[
	\begin{split}
\mathcal{L}_\xi\overline{\Gamma^a_{bc}}&=\underline{\overline{R^a_{cpb}} \xi^p}_1\\
&- \overline{R^a_{bcp}}\xi^p
-\overline{R^a_{pbc}}\xi^p- \underline{\overline{R^a_{cpb}}\xi^p}_1\\
&+\underline{T^a_{bc:<p>}\xi^p}_3
+T^a_{pb;<c>}\xi^p+ \underline{T^a_{cp;<b>}\xi^p}_2\\
& -\xi^a_{;<e>}T^e_{cb}
-\xi^e_{;<c>}T^a_{be}+\xi^a_{;<cb>}\\
&-\underline{T^a_{cp;<b>} \xi^p}_2
-\underline{T^a_{bc;<p>} \xi^p}_3
	\end{split}
	\]
	\begin{equation}
	\label{eq: Lie derivative of connection,9}
\mathcal{L}_\xi\overline{\Gamma^a_{bc}}=-\overline{R^a_{bcp}}\xi^p
-\overline{R^a_{pbc}}\xi^p
-T^a_{bp;<c>}\xi^p
-\xi^a_{;<e>}T^e_{cb}
-\xi^e_{;<c>}T^a_{be}+\xi^a_{;<cb>}
	\end{equation}
We substitute \eqref{eq: commutator second derivative of vector} into \eqref{eq: Lie derivative of connection,9}
	\begin{equation}
	\label{eq: Lie derivative of connection,10}
\mathcal{L}_\xi\overline{\Gamma^a_{bc}}=-\overline{R^a_{bcp}}\xi^p
-T^a_{bp;<c>}\xi^p
-\underline{T^e_{cb}\xi^a_{;<e>}}_1
-T^a_{be}\xi^e_{;<c>}
-\underline{T^p_{bc}\xi^a_{;<p>}}_1+\xi^a_{;<bc>}
	\end{equation}
\eqref{eq: Lie derivative of connection} follows from \eqref{eq: Lie derivative of connection,10}.
		\end{proof}
		\begin{corollary}
Lie derivative of connection in Rieman space has form
	\begin{equation}
	\label{eq: Lie derivative of connection in Rieman space}
\mathcal{L}_\xi\Gamma^a_{bc}=-R^a_{cbp} \xi^p+\xi^a_{;cb}\\
	\end{equation}
		\end{corollary}
		\begin{proof}
\eqref{eq: Lie derivative of connection in Rieman space} follows
from \eqref{eq: Lie derivative of connection} when $T^a_{bc}=0$
		\end{proof}

			\section{Bianchi Identity}
			\label{section: Bianchi Identity}

		\begin{theorem}
The first Bianchi identity for the space with torsion has form
	\begin{equation}
	\label{eq: Bianchi 1}
	\begin{split}
&T^k_{ij;<m>}+T^k_{mi;<j>}+T^k_{jm;<i>}
+ T^k_{pi}T^p_{jm}+ T^k_{pm}T^p_{ij}+ T^k_{pj}T^p_{mi}\\
&=\overline{R^k_{jmi}}+\overline{R^k_{ijm}}+\overline{R^k_{mij}}
	\end{split}
	\end{equation}
		\end{theorem}
		\begin{proof}
Differential of equation \eqref{eq: Torsion 1} has form
\[
T^k_{ij,m}\theta^m\wedge\theta^i\wedge\theta^j
=(\overline{\Gamma^k_{ji,m}}-\overline{\Gamma^k_{ij,m}})\theta^m\wedge\theta^i\wedge\theta^j
\]
Two forms are equal when their coefficients are equal. Therefore
\[
T^k_{ij,m}+T^k_{mi,j}+T^k_{jm,i}
=\overline{\Gamma^k_{ji,m}}-\overline{\Gamma^k_{ij,m}}
+\overline{\Gamma^k_{im,j}}-\overline{\Gamma^k_{mi,j}}
+\overline{\Gamma^k_{mj,i}}-\overline{\Gamma^k_{jm,i}}
\]
We express derivatives using covariant derivatives
and change order of terms
	\begin{align*}
&T^k_{ij;<m>}-\underline{\overline{\Gamma^k_{pm}} T^p_{ij}}_4
+\underline{\overline{\Gamma^p_{im}} T^k_{pj}}_{2:T}-\underline{\overline{\Gamma^p_{jm}} T^k_{pi}}_{3:T}\\
&+T^k_{mi;<j>}-\underline{\overline{\Gamma^k_{pj}} T^p_{mi}}_5
+\underline{\overline{\Gamma^p_{mj}} T^k_{pi}}_3-\underline{\overline{\Gamma^p_{ij}} T^k_{pm}}_{1:T}\\
&+T^k_{jm;<i>}-\underline{\overline{\Gamma^k_{pi}} T^p_{jm}}_6
+\underline{\overline{\Gamma^p_{ji}} T^k_{pm}}_1-\underline{\overline{\Gamma^p_{mi}} T^k_{pj}}_2\\
&=\overline{\Gamma^k_{ji,m}}-\overline{\Gamma^k_{jm,i}}
+\overline{\Gamma^k_{pm}} \overline{\Gamma^p_{ji}}
-\overline{\Gamma^k_{pi}} \overline{\Gamma^p_{jm}}
-\underline{\overline{\Gamma^k_{pm}} \overline{\Gamma^p_{ji}}}_4
+\underline{\overline{\Gamma^k_{pi}} \overline{\Gamma^p_{jm}}}_6\\
&+\overline{\Gamma^k_{im,j}}-\overline{\Gamma^k_{ij,m}}
+\overline{\Gamma^k_{pj}} \overline{\Gamma^p_{im}}
-\overline{\Gamma^k_{pm}} \overline{\Gamma^p_{ij}}
-\underline{\overline{\Gamma^k_{pj}} \overline{\Gamma^p_{im}}}_5
+\underline{\overline{\Gamma^k_{pm}} \overline{\Gamma^p_{ij}}}_4\\
&+\overline{\Gamma^k_{mj,i}}-\overline{\Gamma^k_{mi,j}}
+\overline{\Gamma^k_{pi}} \overline{\Gamma^p_{mj}}-\overline{\Gamma^k_{pj}} \overline{\Gamma^p_{mi}}
-\underline{\overline{\Gamma^k_{pi}} \overline{\Gamma^p_{mj}}}_6
+\underline{\overline{\Gamma^k_{pj}} \overline{\Gamma^p_{mi}}}_5
	\end{align*}
	\begin{equation}
	\label{eq: Bianchi 1,1}
	\begin{split}
&T^k_{ij;<m>}
+T^p_{mi} T^k_{pj}+T^p_{jm} T^k_{pi}
+T^k_{mi;<j>}
+T^p_{ij} T^k_{pm}
+T^k_{jm;<i>}\\
&=\overline{R^k_{jmi}}
+\overline{R^k_{ijm}}
+\overline{R^k_{mij}}
	\end{split}
	\end{equation}
\eqref{eq: Bianchi 1} follows from \eqref{eq: Bianchi 1,1}.
		\end{proof}


If we get a derivative of form \xEqRef{0405.027}{eq: Curvature}
we will see that the second Bianchi identity does not
depend on the torsion. 

			\section{Killing Vector}
			\label{section: Killing Vector}

Invariance of the metric tensor $g$ under the infinitesimal coordinate transformation
\eqref{eq: infinitesimal displacement}
leads to the \AddIndex{Killing equation}{Killing equation}.

		\begin{theorem}
Killing equation in the metric\hyph affine manifold has form
	\begin{equation}
	\label{eq: Killing equation 1}
\xi^k_{;<a>}g_{kb}+\xi^k_{;<b>}g_{ka}+T^l_{ka}g_{lb}\xi^k+T^l_{kb}g_{la}\xi^k+g_{ab;<k>}\xi^k=0
	\end{equation}
		\end{theorem}
		\begin{proof}
Invariance of the metric tensor $g$ means that its Lie derivative equal $0$
	\begin{equation}
\mathcal{L}_\xi g_{ab}=0
	\label{eq: Killing equation 1,1}
	\end{equation}
\eqref{eq: Killing equation 1} folows from \eqref{eq: Killing equation 1,1} and \eqref{eq: Lie derivative of metric}.
		\end{proof}


		\begin{theorem}
The condition of invariance of the connection in the metric\hyph affine manifold has form
	\begin{equation}
	\label{eq: Killing equation 2}
	\begin{split}
\xi^a_{;<bc>}=\overline{R^a_{bcp}}\xi^p
+T^a_{bp;<c>}\xi^p
+T^a_{bp}\xi^p_{;<c>}
	\end{split}
	\end{equation}
		\end{theorem}
		\begin{proof}
Because connection is invariant under the infinitesimal transformation we have
	\begin{equation}
	\label{eq: Killing equation 2,1}
\mathcal{L}_\xi\overline{\Gamma^a_{bc}}=0
	\end{equation}
\eqref{eq: Killing equation 2} follows from \eqref{eq: Killing equation 2,1}
and \eqref{eq: Lie derivative of connection}.
		\end{proof}

We call equation \eqref{eq: Killing equation 2}
the \AddIndex{Killing equation of second type}{Killing equation second type}
and vector $\xi^a$
\AddIndex{Killing vector of second type}{Killing vector second type}.

		\begin{theorem}
Killing vector of second type satisfies equation
	\begin{equation}
	\label{eq: Killing equation 2, colorary}
	\begin{split}
0&=\overline{R^a_{bcp}}\xi^p+\overline{R^a_{cpb}}\xi^p+\overline{R^a_{pbc}}\xi^p
\\ &+T^a_{bp;c}\xi^p+T^a_{pc;b}\xi^p
+T^p_{cb}\xi^a_{;<p>}
+T^a_{bp}\xi^p_{;<c>}
+T^a_{pc}\xi^p_{;<b>}
	\end{split}
	\end{equation}
		\end{theorem}
		\begin{proof}
From \eqref{eq: Killing equation 2}
and \eqref{eq: commutator second derivative of vector}
it follows that
	\begin{equation}
	\label{eq: Killing equation 2,4}
	\begin{split}
\overline{R^a_{p bc}}\xi^p
-T^p_{bc}\xi^a_{;<p>}
&=\overline{R^a_{cbp}}\xi^p
+T^a_{cp;<b>}\xi^p
+T^a_{cp}\xi^p_{;<b>}\\
&-\overline{R^a_{bcp}}\xi^p
-T^a_{bp;<c>}\xi^p
-T^a_{bp}\xi^p_{;<c>}
	\end{split}
	\end{equation}
\eqref{eq: Killing equation 2, colorary} follows from \eqref{eq: Killing equation 2,4}.
		\end{proof}

		\begin{corollary}
The Killing equation of second type in the Riemann space is the identity.
The connection in the Riemann space is invariant under
any infinitesimal transformation \eqref{eq: infinitesimal displacement} 
		\end{corollary}
		\begin{proof}
First of all the torsion is 0. The rest is the consequence of the first Bianchi identity.
		\end{proof}

			\section{Cartan Transport}
			\label{section: Cartan Transport}

Theorems \ref{theorem: xtremeLine} and \ref{theorem: FrenetTransfer} state that
the movement along a line causes an additional
to the parallel transport transformation of a vector. This transformation is very important
and we call it the \AddIndex{Cartan transport}{Cartan transport}.
We introduce the \AddIndex{Cartan symbol}{Cartan symbol}
\symb{\Gamma(C)^i_{kl}}0{Cartan symbol}
\[\ShowSymbol{Cartan symbol}=\frac 1 2 g^{im}(g_{kl;m}-g_{km;l}-g_{ml;k})\]
and the \AddIndex{Cartan connection}{Cartan connection}
\symb{\overbrace{\Gamma^i_{kl}}}0{overbrace Gamma i kl}
\[\ShowSymbol{overbrace Gamma i kl}
=\Gamma^i_{kl}-\Gamma(C)^i_{kl}
=\Gamma^i_{kl}-\frac 1 2 g^{im}(g_{kl;m}-g_{km;l}-g_{ml;k})\]
Using the Cartan connection we can write the Cartan transport as
\[da^i=-\overbrace{\Gamma^i_{kl}}a^k dx^l\]
Respectively we define the
\AddIndex{Cartan derivative}{Cartan derivative}
\symb{\overbrace{\nabla_l} a^i}0{overbrace nabla_l}
\[\ShowSymbol{overbrace nabla_l}= a^i_{;\{l\}}=\partial_l a^i+\overbrace{\Gamma^i_{kl}}a^k\]
\symb{\overbrace{D} a^i}0{overbrace D}
\[\ShowSymbol{overbrace D}=da^i+\overbrace{\Gamma^i_{kl}}a^k dx^l\]

		\begin{theorem}
		\label{theorem: LengthTangentVector}
The Cartan transport along an extreme
line holds length of the tangent vector.
		\end{theorem}
		\begin{proof}
Let $$v^i=\frac {dx^i}{ds}$$ be the tangent vector to an extreme curve.
From theorem \ref{theorem: xtremeLine} it follows that
$$\frac {Dv^l} {ds} =
\frac 1 2 g^{il} \left(g_{kj;i} - g_{ik;j}
- g_{ij;k}\right) v^k v^j
$$
and
$$\frac {Dg_{kl} v^k v^l}{ds}
= \frac {Dg_{kl}} {ds}v^k v^l
+ g_{kl}\frac {Dv^k} {ds} v^l
+ g_{kl}v^k \frac {Dv^l} {ds}=$$
$$= g_{kl;p}v^p v^k v^l+$$
$$+ g_{kl}g^{ik} \frac 1 2 \left(g_{rj;i} - g_{ir;j}
- g_{ij;r}\right) v^r v^j v^l+$$
$$+ g_{kl}v^k g^{il} \frac 1 2 \left(g_{rj;i} - g_{ir;j}
- g_{ij;r}\right) v^r v^j=$$
$$= g_{kl;p}v^p v^k v^l
+ \left(g_{rj;l} - g_{lr;j}
- g_{lj;r}\right) v^r v^j v^l = 0$$
Therefore the length of the vector $v^i$ does not change along the extreme curve.
		\end{proof}

We extend the Cartan transport to any geometrical object like we do
for the parallel transport.

\begin{theorem}
\[
g_{ij;\{l\}}=0
\]
\end{theorem}

		\begin{proof}
\[
\overbrace{\nabla_l} g_{ij}=\partial_l g_{ij}
-\overbrace{\Gamma^k_{il}}g_{kj}
-\overbrace{\Gamma^k_{jl}}g_{ik}=
\]
\[
=g_{ij;l}
+\frac 1 2 g^{km}(g_{il;m}-g_{im;l}-g_{ml;i})g_{kj}
+\frac 1 2 g^{km}(g_{jl;m}-g_{jm;l}-g_{ml;j})g_{ik}=0
\]
		\end{proof}

The Cartan connection $\overbrace{\Gamma^i_{kl}}$ differs from the connection $\Gamma^i_{kl}$
by additional term which is symmetric tensor.
For any connection we introduce standard way derivative and curvature.
Statemants of geometry and physics
have the same form independently of whether I use the connection $\Gamma^i_{kl}$ or the Cartan connection.
To show this we can generalize the idea of the Cartan connection
and consider connection defined by equation
\symb{\overline{\Gamma^i_{kl}}}0{conection overline}
	\begin{equation}
\ShowSymbol{conection overline}=\Gamma^i_{kl}+A^i_{kl}
	\label{eq: conection overline}
	\end{equation}
where $A$ is $0$, or the Cartan symbol or any other symmetric tensor.
Respectively we define the derivative
\symb{\overline{\nabla_l} a^i}0{overline nabla_l, definition 1}
\symb{a^i_{;<l>}}0{overline nabla_l, definition 2}
\[\ShowSymbol{overline nabla_l, definition 1}=
\ShowSymbol{overline nabla_l, definition 2}=
\partial_l a^i+\overline{\Gamma^i_{kl}}a^k\]
\symb{\overline{D} a^i}0{overline D}
\[\ShowSymbol{overline D}=da^i+\overline{\Gamma^i_{kl}}a^k dx^l\]
and curvature
\symb{\overline{R^a_{bij}}}0{GLn curvature_overline}
	\begin{equation}
\ShowSymbol{GLn curvature_overline}
=\partial_i\overline{\Gamma^a_{bj}}-\partial_j\overline{\Gamma^a_{bi}}
+\overline{\Gamma^a_{ci}}\overline{\Gamma^c_{bj}}
-\overline{\Gamma^a_{cj}}\overline{\Gamma^c_{bi}}
	\label{eq: GLn curvature_overline}
	\end{equation}
This connection has the same torsion
	\begin{equation}
T^a_{cb} =
\overline{\Gamma^a_{bc}}-\overline{\Gamma^a_{cb}}
\label{eq: Torsion coordinates_overline}
	\end{equation}

In this context theorem \ref{theorem: LengthTangentVector} means that extreme line is geodesic line
for the Cartan connection.

		\begin{theorem}
		\label{theorem: curvature_overline}
Curvature of connection \eqref{eq: conection overline} has form
	\begin{equation}
\overline{R^a_{bde}}=R^a_{bde}
+A^a_{be;d}-A^a_{bd;e}+A^a_{cd}A^c_{be}-A^a_{ce}A^c_{bd}+S^p_{de}A^a_{bp}
	\label{curvature_overline}
	\end{equation}
where $R^a_{bde}$ is curvature of connection $\Gamma^i_{kl}$
		\end{theorem}
		\begin{proof}
	\begin{align*}
\overline{R^a_{bde}}&=\overline{\Gamma^a_{be,d}}-\overline{\Gamma^a_{bd,e}}
+\overline{\Gamma^a_{cd}}\ \overline{\Gamma^c_{be}}-\overline{\Gamma^a_{ce}}\ \overline{\Gamma^c_{bd}}\\
&=\Gamma^a_{be,d}+A^a_{be,d}-\Gamma^a_{bd,e}-A^a_{bd,e}\\
&+(\Gamma^a_{cd}+A^a_{cd})(\Gamma^c_{be}+A^c_{be})-(\Gamma^a_{ce}+A^a_{ce})(\Gamma^c_{bd}+A^c_{bd})\\
&=\Gamma^a_{be,d}+A^a_{be,d}-\Gamma^a_{bd,e}-A^a_{bd,e}\\
&+\Gamma^a_{cd}\Gamma^c_{be}+\Gamma^a_{cd}A^c_{be}+A^a_{cd}\Gamma^c_{be}+A^a_{cd}A^c_{be}\\
&-\Gamma^a_{ce}\Gamma^c_{bd}-A^a_{ce}\Gamma^c_{bd}-\Gamma^a_{ce}A^c_{bd}-A^a_{ce}A^c_{bd}\\
&=R^a_{bde}+A^a_{be,d}-A^a_{bd,e}\\
&+\Gamma^a_{cd}A^c_{be}+A^a_{cd}\Gamma^c_{be}+A^a_{cd}A^c_{be}\\
&-A^a_{ce}\Gamma^c_{bd}-\Gamma^a_{ce}A^c_{bd}-A^a_{ce}A^c_{bd}\\
&=R^a_{bde}\\
&+A^a_{be;d}-\underline{\Gamma^a_{pd}A^p_{be}}_2+\underline{\Gamma^p_{bd}A^a_{pe}}_4+\underline{\Gamma^p_{ed}A^a_{bp}}_{1:S}\\
&-A^a_{bd;e}+\underline{\Gamma^a_{pe}A^p_{bd}}_3-\underline{\Gamma^p_{be}A^a_{pd}}_5-\underline{\Gamma^p_{de}A^a_{bp}}_1\\
&+\underline{\Gamma^a_{cd}A^c_{be}}_2+\underline{\Gamma^c_{be}A^a_{cd}}_5+A^a_{cd}A^c_{be}\\
&-\underline{\Gamma^c_{bd}A^a_{ce}}_4-\underline{\Gamma^a_{ce}A^c_{bd}}_3-A^a_{ce}A^c_{bd}
	\end{align*}
		\end{proof}

		\begin{corollary}
\AddIndex{Cartan curvature}{Cartan curvature} has next form
\symb{\overbrace{R^a_{bde}}}0{Cartan curvature}
	\begin{equation}
	\begin{split}
\ShowSymbol{Cartan curvature}&=R^a_{bde}
-\Gamma(C)^a_{be;d}+\Gamma(C)^a_{bd;e}\\
&+\Gamma(C)^a_{cd}\Gamma(C)^c_{be}-\Gamma(C)^a_{ce}\Gamma(C)^c_{bd}-T^p_{de}\Gamma(C)^a_{bp}
	\end{split}
	\label{eq: Cartan curvature}
	\end{equation}
		\end{corollary}

\def\texNewton{}
\ifx\PrintBook\Defined
				\chapter{Metric Affine Gravity}
\fi

\section{Newton's Laws: Scalar Potential}
\label{section: Newton Laws: Scalar Potential}

The knowledge of dynamics of a point particle is important for us
because we can study how the particle interacts with external fields
as well as the properties of the particle itself.

To study the movement of a point particle we can use a potential of a certain field.
The potential may be scalar or vector.

In case of scalar potential we assume that a point particle
has rest mass $m$ and we use lagrangian function in the following form
\[L = - m c ds - U dx^0\]
where $U$ is \AddIndex{scalar potential}{scalar potential}
or \AddIndex{potential energy}{potential energy}.

		\begin{theorem}
		\label{theorem: First Newton law}
(\AddIndex{First Newton law}{First Newton law})
If $U = 0$ (therefore we consider free movement)
a body chooses trajectory with extreme length.
		\end{theorem}

		\begin{theorem}
(\AddIndex{Second Newton law}{Second Newton law})
A trajectory of point particle satisfies the differential equation
	\begin{equation}
\frac {\overbrace{D}u^l} {ds} =
\frac {u^0} {m c} F^l
	\label{eq: Newton}
	\end{equation}
\[u^j = \frac {dx^l} {ds}\]
where we introduced force
	\begin{equation}
F^l = g^{il} \frac{\partial U}{\partial x^i}
\label{eq: potential force}
	\end{equation}
		\end{theorem}
		\begin{proof}

Using \eqref{eq: VarLineLength},
we can write variation of the lagrangian as
$$\frac 1 2 m c \left(g_{kl;i} - g_{ik;l}
- g_{il;k}\right) u^k u^j ds
- m c g_{ij} Du^j
+ \frac {\partial U} {\partial x^i} dx^0 = 0$$
The statement of the theorem follows from this.
		\end{proof}

			\section{Newton's Laws: Vector Potential}
In section \ref{section: Newton Laws: Scalar Potential}
we learned dynamics when potential is scalar.
However in electrodynamics we have \AddIndex{vector potential}
{vector potential} $A^k$.
In this case action is
\[
S = \int^{t_2}_{t_1} \left( - m c ds - \frac e c A_l dx^l \right)
\]
\[
A_c = g_{cd} A^d
\]

		\begin{theorem}
The trajectory of a particle moving in the vector field satisfies the differential equation
\[
\frac {\overbrace{D}u^j} {ds} =
\frac e {m c^2} g^{ij} F_{li} u^l
\]
\[u^j = \frac {dx^l} {ds}\]
where we introduce a \AddIndex{field-strength tensor}
{field-strength tensor}
\[F_{dc} = A_{d;c} - A_{c;d} + S^p_{dc} A_p
 = \overbrace{\nabla_c}A_d - \overbrace{\nabla_d}A_c + S^p_{dc} A_p\]
		\end{theorem}
		\begin{proof}

Using \eqref{eq: VarLineLength}, we can write the variation of the action as
\[
\delta S =\]
\[=\int^{t_2}_{t_1} \left( - m c \left(
\frac 1 2 \left( g_{kj;i}
- g_{ij;k} - g_{ik;j} \right) u^k u^j ds
- g_{ij} D u^j\right) \delta x^i
- \frac e c \left(\delta A_l dx^l + A_l \delta dx^l \right)
\right)
\]
We can estimate the second term like
\[
- \frac e c \left(A_{l,k} dx^l \delta x^k + A_l d\delta x^l \right) =
\]
\[
=- \frac e c \left(A_{l;k} dx^l \delta x^k
+ \Gamma^p_{lk} A_p dx^l \delta x^k
+ A_l d\delta x^l \right) =
\]
\[
=- \frac e c \left(A_{k;l} dx^l \delta x^k +
\left(A_{l;k} - A_{k;l}\right)dx^l \delta x^k+
S^p_{lk} A_p dx^l \delta x^k + \Gamma^p_{kl} A_p dx^l \delta x^k
+ A_l d\delta x^l \right) =
\]
\[
=- \frac e c \left(DA_k \delta x^k + A_k D\delta x^k +
\left(A_{l;k} - A_{k;l}\right)dx^l \delta x^k
+ S^p_{lk} A_p dx^l \delta x^k \right) =
\]
\[
=- \frac e c \left(\underline{d\left(A_k \delta x^k\right)} +
\left( A_{l;k} - A_{k;l}
+ S^p_{lk} A_p \right) dx^l \delta x^k\right)
\]
The integral of the underlined term is $0$ because points, when $t = t_1$ and $t = t_2$, are fixed.
Therefore
\[
- m c \left(
\frac 1 2 \left( g_{kj;i}
- g_{ij;k} - g_{ik;j} \right) u^k u^j ds
- g_{ij} D u^j\right)-
\frac e c F_{li} dx^l = 0
\]
The statement of the theorem follows from this.
		\end{proof}

The dependence of field-strength tensor on derivative of metric follows from
this theorem. It changes form of Einstein equation and momentum of gravitational field
appears in case of vector field.

		\begin{theorem}
A field-strength tensor does not change when vector potential changes like
\[A'_j=A_j+\partial_j\Lambda\]
where $\Lambda$ is an arbitrary function of $x$.
		\end{theorem}
		\begin{proof}
Change in a field-strength tensor is
\[(\partial_d\Lambda){}_{;c} - (\partial_c\Lambda){}_{;d} + S^p_{dc} \partial_p\Lambda =\]
\[\partial_{cd}\Lambda - \Gamma^p_{dc} \partial_p\Lambda
- \partial_{dc}\Lambda+\Gamma^p_{cd} \partial_p\Lambda + S^p_{dc}\partial_p\Lambda =0\]
This proves the theorem.
		\end{proof}

\OpenBiblio


\BiblioItem{texSpaceTime}{Einstein: Geometry and Experience}
{
Einstein, Geometry and Experience, (1921)
}%

\BiblioItem{texGenRelativity}{Ghez}
{
Ghez et al.,
The First Measurement of Spectral Lines in a Short-Period Star Bound to the Galaxy's Central Black Hole: A Paradox of Youth,
\href{http://www.journals.uchicago.edu/ApJ/journal/issues/ApJL/v586n2/16990/brief/16990.abstract.html}{ApJL, 586, L127} (2003),
eprint \href{http://arxiv.org/abs/astro-ph/0302299}{arXiv:astro-ph/0302299} (2003)
}%

\BiblioItem{texGenRelativity}{Schodel}
{
R. Sch\"odel et al.,
A star in a 15.2-year orbit around the supermassive black hole at the centre of the Milky Way,
\href{http://www.nature.com/cgi-taf/DynaPage.taf?file=/nature/journal/v419/n6908/abs/nature01121_fs.html}{Nature 419, 694} (2002)
}%

\BiblioItem{texAffine,texGeomObject}{Mielke}
{
Eckehard W. Mielke, Affine generalization of the Komar complex of general relativity,
\href{http://prola.aps.org/searchabstract/PRD/v63/i4/e044018}{Phys. Rev. D 63, 044018} (2001)
}%

\BiblioItem{texAffine}{Obukhov}
{
Yu. N. Obukhov and J. G. Pereira, Metric\hyph affine approach to teleparallel gravity,
\href{http://scitation.aip.org/getabs/servlet/GetabsServlet?prog=normal&id=PRVDAQ000067000004044016000001&idtype=cvips&gifs=Yes}
{Phys. Rev. D 67, 044016} (2003),
eprint \href{http://arxiv.org/abs/gr-qc/0212080}{arXiv:gr-qc/0212080} (2002)
}%

\BiblioItem{texAffine}{Sardanashvily}
{
Giovanni Giachetta, Gennadi Sardanashvily, Dirac Equation in Gauge and Affine-Metric Gravitation Theories,
eprint \href{http://arxiv.org/abs/gr-qc/9511035}{arXiv:gr-qc/9511035} (1995)
}%

\BiblioItem{texAffine}{Gauge}
{
Frank Gronwald and Friedrich W. Hehl, On the Gauge Aspects of Gravity, eprint
\href{http://arxiv.org/abs/gr-qc/9602013}{arXiv:gr-qc/9602013} (1996)
}%

\BiblioItem{texAffine}{Neeman}
{
Yuval Neeman, Friedrich W. Hehl, Test Matter in a Spacetime with Nonmetricity, eprint
\href{http://arxiv.org/abs/gr-qc/9604047}{arXiv:gr-qc/9604047} (1996)
}%

\BiblioItem{texTidal,texAffine,texGeomObject}{torsion}
{
F. W. Hehl, P. von der Heyde, G. D. Kerlick, and J. M. Nester,
General relativity with spin and torsion: Foundations and prospects,
\href{http://prola.aps.org/abstract/RMP/v48/i3/p393_1}{Rev. Mod. Phys. 48, 393} (1976)
}%

\BiblioItem{texTidal,texNewton}{Megged}
{
O. Megged, Post-Riemannian Merger of Yang-Mills Interactions with Gravity,
eprint \href{http://arxiv.org/abs/hep-th/0008135}{arXiv:hep-th/0008135} (2001)
}%


\BiblioItem{texNewton}{gr-qc-9604027}
{
Yu.N. Obukhov, E.J. Vlachynsky, W. Esser, R. Tresguerres and F.W. Hehl,
An exact solution of the metric\hyph affine gauge theory with dilation, shear, and spin charges,
eprint \href{http://arxiv.org/abs/gr-qc/9604027}{arXiv:gr-qc/9604027} (1996)
}%

\BiblioItem{texLagrange}{Weinberg}
{
Steven Weinberg. The Quantum Theory of Fields. Cambridge university press.
}%

\BiblioItem{texLagrange}{Reinhardt}
{
Greiner Reinhardt. Field Quantization. Springer.
}%

\BiblioItem{texLagrange}{Landau}
{
L. D. Landau, E. M. Lifshich, The classical theory of fields.
Oxford, New York, Pergamon Press
}%

\BiblioItem{texTidal}{Wheeler}
{
Ignazio Ciufolini, John Wheeler. Gravitation and Inertia.
Princeton university press.
}%

\BiblioItem{texPrefaceTidal,texPrefaceRefernceFrame}{Anderson02}
{
J. D. Anderson, P. A. Laing, E. L. Lau, A. S. Liu, M. M. Nieto, and S. G. Turyshev,
Study of the anomalous acceleration of Pioneer 10 and 11,
\href{http://prola.aps.org/searchabstract/PRD/v65/i8/e082004}{Phys. Rev. D 65, 082004, 50 pp.}, (2002),
eprint \href{http://arxiv.org/abs/gr-qc/0104064}{arXiv:gr-qc/0104064} (2001)
}%

\BiblioItem{texTidal}{Anderson98}
{
J. D. Anderson, P. A. Laing, E. L. Lau, A. S. Liu, M. M. Nieto, and S. G. Turyshev,
Indication, from Pioneer 10/11, Galileo, and Ulysses Data, of an Apparent Anomalous, Weak, Long-Range Acceleration,
\href{http://prola.aps.org/abstract/PRL/v81/i14/p2858_1}{Phys. Rev. Lett. 81, 2858}, (1998),
eprint \href{http://arxiv.org/abs/gr-qc/9808081}{arXiv:gr-qc/9808081} (1998)
}%


\BiblioItem{texReferenceFrame,texFiberedAlgebra}{Serge Lang}
{
Serge Lang,
Algebra, Springer, 2002
}%

\BiblioItem{texFiberedAlgebra,texTstarMorphism}{Burris Sankappanavar}
{
S. Burris, H.P. Sankappanavar,
A Course in Universal Algebra, Springer-Verlag (March, 1982),
\\eprint
\href{http://www.math.uwaterloo.ca/~snburris/htdocs/ualg.html}
{http://www.math.uwaterloo.ca/~snburris/htdocs/ualg.html}
\\(The Millennium Edition)
}%

\BiblioItem{texGeomObject}{Shilov}
{
G. E. Shilov,
Calculus, Multivariable Functions,
Moscow, Nauka, 1972
}%

\BiblioItem{texAffine,texRepresentation,texBasis,texDrcBasis,texLinearMap,texVectorSpace,texPolyvector}{Rashevsky}
{
P. K. Rashevsky, Riemann Geometry and Tensor Calculus,\\
Moscow, Nauka, 1967
}%

\BiblioItem{texPolyvector}{Dubrovin Fomenko Novikov part 1}
{
B. A. Dubrovin, A. T. Fomenko, S. P. Novikov,
Modern Geometry - Methods and Applications,\\
Part 1, The Geometry of Surfaces, Transformation Groups, and Fields,\\
Translated by Robert G. Burns,\\
Springer - New York, 1992
}%

\BiblioItem{texDrcBasis,texBasis}{Korn}
{
Granino A. Korn, Theresa M. Korn,
Mathematical Handbook for Scientists and Engineer,
McGraw-Hill Book Company, New York, San Francisco,
Toronto, London, Sydney, 1968
}%

\BiblioItem{texBundle}{Hocking Young Topology}
{
John G. Hocking, Gail S. Young,
Topology,\\
Courier Dover Publications, 1988
}%


\BiblioItem{texPrefaceRefernceFrame}{Tartaglia}
{
Angelo Tartaglia and Matteo Luca Ruggiero,
Angular Momentum Effects in Michelson\Hyph Morley Type Experiments,
Gen.Rel.Grav. 34, 1371-1382 (2002),\\
eprint \href{http://arxiv.org/abs/gr-qc/0110015}{arXiv:gr-qc/0110015} (2001)
}%

\BiblioItem{texPrefaceRefernceFrame}{Tomozawa}
{
Yukio Tomozawa, Speed of Light in Gravitational Fields, eprint
\href{http://arxiv.org/abs/astro-ph/0303047}{arXiv:astro-ph/0303047} (2004)
}%

\BiblioItem{texPrefaceRefernceFrame}{Magueijo}
{
Joao Magueijo,
Covariant and locally Lorentz-invariant varying speed of light theories,
\href{http://prola.aps.org/abstract/PRD/v62/i10/e103521}{Phys. Rev. D 62, 103521} (2000),
eprint \href{http://arxiv.org/abs/gr-qc/0007036}{arXiv:gr-qc/0007036} (2000)
}%

\BiblioItem{texPrefaceRefernceFrame}{Bassett}
{
Bruce A. Bassett, Stefano Liberati, Carmen Molina-Paris, and Matt Visser,
Geometrodynamics of variable-speed-of-light cosmologies,
\href{http://prola.aps.org/abstract/PRD/v62/i10/e103518}{Phys. Rev. D 62}, 103518 (2000),
eprint \href{http://arxiv.org/abs/astro-ph/0001441}{arXiv:astro-ph/0001441} (2000)
}%

\BiblioItem{texPrefaceRefernceFrame}{Straumann}
{
Lochlainn O'Raifeartaigh and Norbert Straumann,
Gauge theory: Historical origins and some modern developments,
\href{http://prola.aps.org/abstract/RMP/v72/i1/p1_1}{Rev. Mod. Phys. 72, 1} (2000)
}%

\BiblioItem{texPrefaceRefernceFrame}{Lammerzahl}
{
Claus L\"ammerzahl, Mark P. Haugan,
On the interpretation of Michelson\Hyph Morley experiments,
{Phys. Lett. A282 223-229} (2001),\\
eprint \href{http://arxiv.org/abs/gr-qc/0103052}{arXiv:gr-qc/0103052} (2001)
}%

\BiblioItem{texPrefaceRefernceFrame}{Muller}
{
Holger Muller et al.,
Modern Michelson-Morley Experiment using Cryogenic Optical Resonators,
\href{http://prola.aps.org/searchabstract/PRL/v91/i2/e020401}{Phys. Rev. Lett. 91, 020401} (2003),
eprint \href{http://arxiv.org/abs/physics/0305117}{arXiv:physics/0305117} (2000)
}%

\BiblioItem{texPrefaceRefernceFrame,texPrefaceTidal}{Ranada}
{
Antonio F. Ranada,
Pioneer acceleration and variation of light speed: experimental situation,
eprint \href{http://arxiv.org/abs/gr-qc/0402120}{arXiv:gr-qc/0402120} (2004)
}%

\BiblioItem{texBiring,texVectorSpace}{math.QA-0208146}
{
I. Gelfand, S. Gelfand, V. Retakh, R. Wilson,
Quasideterminants,\\
eprint \href{http://arxiv.org/abs/math.QA/0208146}{arXiv:math.QA/0208146} (2002)
}%

\BiblioItem{texBiring,texVectorSpace}{q-alg-9705026}
{
I.Gelfand, V.Retakh,
Quasideterminants, I,\\
eprint \href{http://arxiv.org/abs/q-alg/9705026}{arXiv:q-alg/9705026} (1997)
}%

\BiblioItem{texVectorSpace}{Gelfand Retakh 1991}
{
I. Gelfand and V. Retakh, Determinants of Matrices over Noncommutative Rings, Funct.
Anal. Appl. 25 (1991), no. 2, 91-102
}%

\BiblioItem{texVectorSpace}{Gelfand Retakh 1992}
{
I. Gelfand and V. Retakh, A Theory of Noncommutative Determinants and Characteristic
Functions of Graphs, Funct. Anal. Appl. 26 (1992), no. 4, 1-20
}%

\BiblioItem{texVectorSpace}{hep-th-9407124}
{
I. M. Gelfand, D. Krob, A. Lascoux, B. Leclerc, V.S. Retakh and J.-Y. Thibon,
Noncommutative symmetric functions,\\
eprint \href{http://arxiv.org/abs/hep-th/9407124}{arXiv:hep-th/9407124} (1994)
}%

\BiblioItem{texVectorSpace}{Carl Faith 1}
{
Carl Faith, Algebra: Rings, Modules and Categories I,
Springer - Verlag, Berlin - Heidelberg - New York, 1973
}%



\BiblioItem{texDrcReferenceFrame,texRefernceFrame,texLie,texLieRepresentation}{0412.391}
{
Aleks Kleyn,
Basis Manifold,
eprint \href{http://arxiv.org/abs/math.DG/0412391}{arXiv:math.DG/0412391} (2004)
}%

\BiblioItem{texAffine}{0405.027}
{
Aleks Kleyn,
Reference Frame in General Relativity,\\
eprint \href{http://arxiv.org/abs/gr-qc/0405027}{arXiv:gr-qc/0405027} (2004)
}%

\BiblioItem{texTidal}{0405.028}
{
Aleks Kleyn, Metric\hyph Affine Manifold,
eprint \href{http://arxiv.org/abs/gr-qc/0405028}{arXiv:gr-qc/0405028} (2004)
}%

\BiblioItem{texFiberedAlgebra,texBundleRelation,texTstarMorphism}{0701.238}
{
Aleks Kleyn,
Lectures on Linear Algebra over Skew Field,\\
eprint \href{http://arxiv.org/abs/math.GM/0701238}{arXiv:math.GM/0701238} (2007)
}%

\BiblioItem{texBundleRelation,texPrefaceRelation}{0702.561}
{
Aleks Kleyn,
Algebra Bundle,\\
eprint \href{http://arxiv.org/abs/math.DG/0702561}{arXiv:math.DG/0702561} (2007)
}%


\BiblioItem{texPolymodule}{math.RA-0501237v1}
{
Aleks Kleyn,
Module Over Skew-Field, version 1,\\
eprint \href{http://arxiv.org/abs/math/0501237v1}{arXiv:math.RA/0501237v1} (2005)
}%

\ifx\texBiring\Defined
\else
\BiblioItem{texVectorSpace,texFiberedAlgebra}{0612.111}
{
Aleks Kleyn,
Biring of Matrices,\\
eprint \href{http://arxiv.org/abs/math.OA/0612111}{arXiv:math.OA/0612111} (2006)
}%
\fi

\ifx\texBundleRelation\Defined
\else
\BiblioItem{texFiberedMorphism}{0707.2246}
{
Aleks Kleyn,
Fibered Correspondence,\\
eprint \href{http://arxiv.org/abs/0707.2246}{arXiv:0707.2246} (2007)
}%
\fi

\ifx\PrintBook\Defined
\BiblioItem{texPrefaceRelation}{0707.2246}
{
Aleks Kleyn,
Fibered Correspondence,\\
eprint \href{http://arxiv.org/abs/0707.2246}{arXiv:0707.2246} (2007)
}%
\fi

\BiblioItem{texHomotopy}{q-alg-9705009}
{
John C. Baez,
An Introduction to n-Categories,\\
eprint \href{http://arxiv.org/abs/q-alg/9705009}{arXiv:q-alg/9705009} (1997)
}%

\BiblioItem{texSpaceTime}{Einstein: Isaak Newton}
{
Albert Einstein,
Isaak Newton,
Manchester Guardian, 16, 234 - 235 (1927)
}%

\BiblioItem{texPrefaceRelation}{Tolstoi about Anna Karenina}
{
Tolstoi about Anna Karenina,
in book A Karenina Companion, by C. J. G. Turner,
published by Wilfrid Laurier University Press (August 1993)
}%

\BiblioItem{texBundleRelation,texPrefaceRelation,texTstarMorphism,texBundle}
{Cohn: Universal Algebra}
{
Paul M. Cohn,
Universal Algebra,
Springer, 1981
}%

\BiblioItem{texBundle}
{Maunder: Algebraic Topology}
{
C. R. F. Maunder,
Algebraic Topology,
Dover Publications, Inc, Mineola, New York, 1996
}%

\BiblioItem{texFiberedAlgebra}{Pommaret: Partial Differential Equations}
{
J.-F. Pommaret,
Partial Differential Equations and Group Theory,
Springer, 1994
}%

\BiblioItem{texBundleRelation}{Bourbaki: Set Theory}
{
N. Bourbaki,
Theory of sets,
Springer, 2004
}%

\BiblioItem{texBundle,texCartesian,texFiberedAlgebra,texBundleRelation,texFiberedMorphism}
{Bourbaki: General Topology 1}
{
N. Bourbaki,
General Topology, Chapters 1 - 4,
Springer, 1989
}

\BiblioItem{texCalculus}{Bourbaki: General Topology: Chapter 5 - 10}
{
N. Bourbaki,
General Topology, Chapters 5 - 10,
Springer, 1989
}

\BiblioItem{texCalculus}{Bourbaki: Topological Vector Space}
{
N. Bourbaki,
Topological Vector Spaces, Chapters 1 - 5,
Springer, 1987
}

\BiblioItem{texCalculus}{Pontryagin: Topological Group}
{
L. S. Pontryagin,
Selected Works, Volume Two, Topological Groups,
Gordon and Breach Science Publishers, 1986
}

\BiblioItem{texFiberedMorphism}{Postnikov: Differential Geometry}
{
Postnikov M. M.,
Geometry IV: Differential geometry,
Moscow, Nauka, 1983
}

\BiblioItem{texFiberedAlgebra,texFiberedMorphism}{Hatcher: Algebraic Topology}
{
Allen Hatcher,
Algebraic Topology,
Cambridge University Press, 2002
}

\BiblioItem{texFiberedMorphism}{geometry of differential equations}
{
Vinogradov, A. M., Krasil'shchik, I. S., and Lychagin, V. V.,
Introduction to geometry of nonlinear differential equations,
Nauka, Moscow, 1986
}

\BiblioItem{texFiberedMorphism}{cohomological analysis}
{
A. M. Vinogradov,
Cohomological Analysis of Partial Differential Equations
and Secondary Calculus,
American Mathematical Society, 2001
}

\BiblioItem{texPolyvector}{0801.1734}
{
Brandon S. DiNunno, Richard A. Matzner,
The Volume Inside a Black Hole,\\
eprint \href{http://arxiv.org/abs/0801.1734v1}{arXiv:0801.1734v1} (2008)
}

\CloseBiblio

\OpenIndex
\SetIndexSpace%
\Index{texLinearMap}
   {$1$-\drc form}%
   {1-drc form, vector spaces}%
\SetIndexSpace%
\Index{texPolymodule}
   {$(2)$\hyph vector space}%
   {(2)-vector space}%
\Index{texBundleRelation}
   {$2$\Hyph ary fibered relation}%
   {2 ary fibered relation}%
\SetIndexSpace%
\Index{texCalculus}
   {$A$\Hyph valued function}%
   {A valued function}%
\Index{texBiring}
   {$(^{\gi a}_{\gi b})$\hyph \CR quasideterminant}%
   {a b cr-quasideterminant}%
\Index{texBiring}
   {$(^{\gi a}_{\gi b})$\hyph \RC quasideterminant}%
   {a b RC-quasideterminant}%
\Index{texCalculus}
   {absolute value on skew field}%
   {absolute value on skew field}%
\Index{texBasis}
   {active representation}%
   {active representation}%
\Index{texDrcBasis}
   {active \sT representation}%
   {active representation, vector space}%
\Index{texBasis}
   {active transformation on basis manifold}%
   {active transformation}%
\Index{texBasis}
   {affine basis}%
   {Affine Basis}%
\Index{texBasis}
   {affine transformation group}%
   {AffineTransformationGroup}%
\Index{texTypeBasis}
   {affine transformation group}%
   {AffineTransformationGroup}%
\Index{texBasis}
   {affine transformation on basis manifold}%
   {affine transformation}%
\Index{texBiring}
   {alternative representation of matrix}%
   {Alternative representation}%
\Index{texRefernceFrame}
   {anholonomic coordinate}%
   {anholonomic coordinate}%
\Index{texRefernceFrame}
   {anholonomic coordinates of connection}%
   {anholonomic coordinates of connection}%
\Index{texRefernceFrame}
   {anholonomic coordinates of vector}%
   {vector anholonomic coordinates}%
\Index{texRefernceFrame}
   {anholonomic coordinates on manifold}%
   {anholonomic coordinates on manifold}%
\Index{texRefernceFrame}
   {anholonomity object}%
   {anholonomity object}%
\Index{texFiberedGroup}
   {antihomomorphism of fibered groups}%
   {antihomomorphism of fibered groups}%
\Index{texBundleRelation}
   {antisymmetric $2$\Hyph ary fibered relation}%
   {antisymmetric 2 ary fibered relation}%
\Index{texFiberedAlgebra}
   {arity of operation}%
   {arity of operation}%
\Index{texTstarRepresentation}
   {associative law for covariant \Ts representation}%
   {associative law for Tstar covariant representation}%
\Index{texDrcMorphism}
   {associative law for \drc linear maps of vector spaces}%
   {associative law for drc linear maps of vector spaces}%
\Index{texVectorSpace}
   {associative law for \Ts vector space}%
   {associative law, Tstar vector space}%
\Index{texLinearMap}
   {associative law for twin representations}%
   {associative law for twin representations}%
\Index{texBundleRelation}
   {associative law of composition of fibered correspondences}%
   {associative law, composition of fibered correspondences}%
\Index{texAffine}
   {auto parallel line}%
   {auto parallel line}%
\SetIndexSpace%
\Index{texBundleRelation}
   {base of fibered correspondence}%
   {base of fibered correspondence}%
\Index{texBundle}
   {base of map}%
   {base of map}%
\Index{texTypeBasis}
   {basis}%
   {}%
\SubIndex{texTypeBasis}
   {affine}%
   {Affine Basis}%
\SubIndex{texTypeBasis}
   {central affine}%
   {Central Affine Basis}%
\SubIndex{texTypeBasis}
   {orthonornal}%
   {Orthonornal Basis}%
\Index{texDrcBasis}
   {basis manifold}%
   {}%
\SubIndex{texTypeBasis}
   {of affine space}%
   {Basis Manifold, Affine Space}%
\SubIndex{texTypeBasis}
   {of central affine space}%
   {Basis Manifold, Central Affine Space}%
\SubIndex{texDrcBasis}
   {of \drc vector space}%
   {basis manifold of vector space}%
\SubIndex{texTypeBasis}
   {of Euclid space}%
   {Basis Manifold, Euclid Space}%
\Index{texBasis}
   {basis manifold of affine space}%
   {Basis Manifold, Affine Space}%
\Index{texBasis}
   {basis manifold of central affine space}%
   {Basis Manifold, Central Affine Space}%
\Index{texBasis}
   {basis manifold of Euclid space}%
   {Basis Manifold, Euclid Space}%
\Index{texBasis}
   {basis manifold of vector space}%
   {basis manifold of vector space}%
\Index{texBasis}
   {basis of vector space}%
   {Basis}%
\Index{texLieRepresentation}
   {basis vector}%
   {}%
\SubIndex{texLieRepresentation}
   {of \sT representation}%
   {basis vector of starT representation}%
\SubIndex{texLieRepresentation}
   {of \Ts representation}%
   {basis vector of Tstar representation}%
\Index{texBiring}
   {biring}%
   {biring}%
\Index{texFiberedMorphism}
   {bundle of level $2$}%
   {bundle of level 2}%
\Index{texFiberedMorphism}
   {bundle of level $n$}%
   {bundle of level n}%
\SetIndexSpace%
\Index{texVectorSpace}
   {\subs rows \dcr vector space}%
   {subs rows dcr vector space}%
\Index{texBiring}
   {\subs row of matrix}%
   {c row}%
\Index{texBiring}
   {$c$\hyph row of matrix}%
   {c-row}%
\Index{texAffine}
   {Cartan connection}%
   {Cartan connection}%
\Index{texAffine}
   {Cartan curvature}%
   {Cartan curvature}%
\Index{texAffine}
   {Cartan derivative}%
   {Cartan derivative}%
\Index{texAffine}
   {Cartan symbol}%
   {Cartan symbol}%
\Index{texAffine}
   {Cartan transport}%
   {Cartan transport}%
\Index{texBundle}
   {Cartesian power $\mathcal{A}$ of bundle $\mathcal{B}$}%
   {Cartesian power of bundle}%
\Index{texBundle}
   {Cartesian power $A$ of set $B$}%
   {Cartesian power of set}%
\Index{texCartesian}
   {Cartesian power of bundle}%
   {Cartesian power of bundle}%
\Index{texCartesian}
   {direct product of bundles}%
   {Cartesian product of bundles}%
\Index{texCartesian}
   {direct product of total spaces}%
   {Cartesian product of total spaces}%
\Index{texHomotopy}
   {category of \drc vector spaces}%
   {category of drc vector spaces}%
\Index{texBundleRelation}
   {category of fibered correspondences over diagonal}%
   {category of fibered correspondences over diagonal}%
\Index{texBundleRelation}
   {category of reduced fibered correspondences}%
   {category of reduced fibered correspondences}%
\Index{texBasis}
   {central affine basis}%
   {Central Affine Basis}%
\Index{texRepresentation}
   {column vector}%
   {column vector}%
\Index{texBundleRelation}
   {commutative diagram of correspondences}%
   {commutative diagram of correspondences}%
\Index{texBundle}
   {compact\hyph open topology}%
   {compact open topology}%
\Index{texDiffEq}
   {completely integrable system}%
   {completely integrable system}%
\Index{texBundleRelation}
   {composition of fibered correspondences}%
   {composition of fibered correspondences}%
\SubIndex{texVectorSpace}
   {of \drc linear equations}%
   {extended matrix, system of drc linear equations}%
\SubIndex{texVectorSpace}
   {of \rcd linear equations}%
   {extended matrix, system of rcd linear equations}%
\Index{texBundleRelation}
   {composition of reduced fibered correspondences}%
   {composition of reduced fibered correspondences}%
\Index{texBiring}
   {condition of reducibility of products}%
   {condition of reducibility of products}%
\Index{texBundleRelation}
   {continuous correspondence}%
   {continuous correspondence}%
\Index{texFiberedGroup}
   {contravariant \Ts representation of fibered group}%
   {Tstar contravariant representation of fibered group}%
\Index{texBasis}
   {coordinate isomorphism}%
   {coordinate isomorphism}%
\Index{texVectorSpace}
   {coordinate isomorphism}%
   {coordinate isomorphism}%
\Index{texVectorSpace}
   {coordinate matrix}%
   {}%
\SubIndex{texVectorSpace}
   {of set of vectors in \dcr rows vector space}%
   {coordinate matrix of set of vectors, dcr vector space}%
\SubIndex{texVectorSpace}
   {of set of vectors in \drc rows vector space}%
   {coordinate matrix of set of vectors, drc vector space}%
\SubIndex{texVectorSpace}
   {of vector in \drc basis}%
   {coordinate matrix of vector in drc basis}%
\Index{texRefernceFrame}
   {coordinate reference frame}%
   {coordinate reference frame}%
\Index{texDrcBasis}
   {coordinate representation in \drc vector space}%
   {coordinate representation, vector space}%
\Index{texBasis}
   {coordinate representation of group in vector space}%
   {coordinate representation, vector space}%
\Index{texBasis}
   {coordinate vector space}%
   {coordinate vector space}%
\Index{texBasis}
   {coordinates of geometrical object}%
   {coordinates of geometrical object, vector space}%
\Index{texDrcBasis}
   {coordinates of geometrical object}%
   {}%
\SubIndex{texDrcBasis}
   {in coordinate vector space}%
   {coordinates of geometrical object, coordinate vector space}%
\SubIndex{texDrcBasis}
   {in vector space}%
   {coordinates of geometrical object, vector space}%
\Index{texBasis}
   {coordinates of geometrical object in coordinate representation}%
   {coordinates of geometrical object, coordinate vector space}%
\Index{texDrcBasis}
   {coordinates of representation}%
   {coordinates of representation}%
\Index{texBasis}
   {coordinates of representation}%
   {coordinates of representation}%
\Index{texVectorSpace}
   {coordinates of set of vectors in \dcr vector space}%
   {coordinates of set of vectors, dcr vector space}%
\Index{texVectorSpace}
   {coordinates of set of vectors in \drc vector space}%
   {coordinates of set of vectors, drc vector space}%
\Index{texVectorSpace}
   {coordinates of vector in \drc basis}%
   {coordinates of vector in drc basis}%
\Index{texBundleRelation}
   {correspondence continuous on the set}%
   {correspondence continuous on the set}%
\Index{texBundleRelation}
   {correspondence of homomorphism}%
   {correspondence of homomorphism}%
\Index{texFiberedGroup}
   {covariant \Ts representation of fibered group}%
   {Tstar covariant representation of fibered group}%
\Index{texBiring}
   {\CR inverse element of biring}%
   {cr-inverse element}%
\Index{texVectorSpace}
   {\CR matrix group}%
   {cr-matrix group}%
\Index{texBiring}
   {\CR power}%
   {cr power}%
\Index{texBiring}
   {\CR product of matrices}%
   {cr-product of matrices}%
\Index{texVectorSpace}
   {\crd vector space}%
   {crd vector space}%
\SetIndexSpace%
\Index{texCalculus}
   {$D$\Hyph valued variable}%
   {D valued variable}%
\Index{texVectorSpace}
   {\dcr basis of \subs rows vector space}%
   {dcr basis, c rows vector space}%
\Index{texVectorSpace}
   {\dcr vector}%
   {dcr vector}%
\Index{texVectorSpace}
   {\dcr vector space}%
   {dcr vector space}%
\Index{texBiring}
   {determinant of matrix}%
   {determinant}%
\Index{texTidal}
   {deviation of trajectories}%
   {deviation of trajectories}%
\Index{texBundleRelation}
   {diagonal in bundle}%
   {diagonal in bundle}%
\Index{texBundleRelation}
   {diagram of correspondences}%
   {diagram of correspondences}%
\Index{texCalculus}
   {differentiable functions of \drc vector space to skew field $D$}%
   {differentiable functions, drc vector space to skew field}%
\Index{texCalculus}
   {differential of mapping of normed \drc vector space to valued skew field}%
   {differential, drc vector space to skew field}%
\Index{texVectorSpace}
   {dimension of \drc vector space}%
   {dimension of vector space}%
\Index{texFiberedGroup}
   {direct product of representations of fibered group}%
   {direct product of representations of fibered group}%
\Index{texRepresentation}
   {direct product of representations of group}%
   {direct product of representations of group}%
\Index{texTstarRepresentation}
   {direct product of \Ts representations of group}%
   {direct product of representations of group}%
\Index{texLieRepresentation}
   {direct sum of representations}%
   {direct sum of representations}%
\Index{texVectorSpace}
   {distributive law}%
   {}%
\SubIndex{texVectorSpace}
   {\Ts vector space}%
   {distributive law, Tstar vector space}%
\Index{texVectorSpace}
   {\drc automorphism of vector space}%
   {automorphism of vector space}%
\Index{texVectorSpace}
   {\drc basis}%
   {}%
\SubIndex{texVectorSpace}
   {for \sups rows vector space}%
   {drc basis, r rows vector space}%
\SubIndex{texVectorSpace}
   {for vector space}%
   {drc basis, vector space}%
\Index{texVectorSpace}
   {\drc coordinate vector space}%
   {drc coordinate vector space}%
\Index{texVectorSpace}
   {\drc isomorphism of vector spaces}%
   {isomorphism of vector spaces}%
\Index{texDrcMorphism}
   {\drc linear map of vector spaces}%
   {drc linear map of vector spaces}%
\Index{texVectorSpace}
   {\drc linear span in vector space}%
   {linear span, vector space}%
\Index{texVectorSpace}
   {\drc linearly dependent vectors}%
   {linearly dependent, vector space}%
\Index{texVectorSpace}
   {\drc linearly independent vectors}%
   {linearly independent, vector space}%
\Index{texVectorSpace}
   {\drc system of linear equations}%
   {system of linear equations}%
\Index{texVectorSpace}
   {\drc vector}%
   {drc vector}%
\Index{texCalculus}
   {\drc vector function}%
   {drc vector function}%
\Index{texVectorSpace}
   {\drc vector space}%
   {drc vector space}%
\Index{texVectorSpace}
   {$D\star$\hyph  vector space}%
   {Dstar vector space}%
\Index{texVectorSpace}
   {$D\star$\hyph product of vector over scalar}%
   {Dstar product of vector over scalar, vector space}%
\Index{texBiring}
   {duality principle for biring}%
   {duality principle for biring}%
\Index{texBiring}
   {duality principle for biring of matrices}%
   {duality principle for biring of matrices}%
\SetIndexSpace%
\Index{texTstarMorphism}
   {effective representation of $\mathfrak{F}$\Hyph algebra $A$}%
   {effective representation of algebra}%
\Index{texFiberedAlgebra}
   {effective representation of fibered $\mathfrak{F}$\Hyph algebra}%
   {effective representation of fibered F-algebra}%
\Index{texFiberedGroup}
   {effective \Ts representation of fibered group}%
   {effective representation of fibered group}%
\Index{texRepresentation}
   {effective representation of group}%
   {effective representation of group}%
\Index{texVectorSpace}
   {effective representation of skew field}%
   {effective representation of skew field}%
\Index{texELie}
   {enhanced Lie group}%
   {enhanced Lie group}%
\Index{texDiffEq}
   {essential parameters}%
   {essential parameters}%
\Index{texBundleRelation}
   {extension of correspondence}%
   {extension of correspondence}%
\Index{texAffine}
   {extreme line}%
   {extreme line}%
\SetIndexSpace%
\Index{texBundleRelation}
   {fibered correspondence from $\mathcal{A}$ to $\mathcal{B}$}%
   {fibered correspondence from A to B}%
\Index{texBundleRelation}
   {fibered correspondence in $\mathcal{A}$}%
   {fibered correspondence in A}%
\Index{texBundleRelation}
   {fibered correspondence of homomorphism}%
   {fibered correspondence of homomorphism}%
\Index{texBundleRelation}
   {fibered equivalence}%
   {fibered equivalence}%
\Index{texFiberedAlgebra}
   {fibered $\mathfrak{F}$\Hyph algebra}%
   {fibered F-algebra}%
\Index{texFiberedAlgebra}
   {fibered $\mathfrak{F}$\Hyph subalgebra}%
   {fibered F-subalgebra}%
\Index{texFiberedAlgebra}
   {fibered group}%
   {fibered group}%
\Index{texFiberedMorphism}
   {fibered identification morphism}%
   {fibered identification morphism}%
\Index{texFiberedMorphism}
   {fibered little group}%
   {fibered little group}%
\Index{texBundle}
   {fibered morphism from bundle $\mathcal{A}$ into $\mathcal{B}$}%
   {fibered morphism from A into B}%
\Index{texFiberedMorphism}
   {fibered natural morphism}%
   {fibered natural morphism}%
\Index{texBundleRelation}
   {fibered ordering}%
   {fibered ordering}%
\Index{texBundleRelation}
   {fibered preordering}%
   {fibered preordering}%
\Index{texFiberedAlgebra}
   {fibered ring}%
   {fibered ring}%
\Index{texFiberedMorphism}
   {fibered stability group}%
   {fibered stability group}%
\Index{texBundle}
   {fibered subset}%
   {fibered subset}%
\Index{texNewton}
   {field-strength tensor}%
   {field-strength tensor}%
\Index{texBundleRelation}
   {filter $\mathfrak{F}$ converges to $A$}%
   {filter converges}%
\Index{texNewton}
   {first Newton law}%
   {First Newton law}%
\Index{texFiberedMorphism}
   {free \Ts representation of fibered group}%
   {free representation of fibered group}%
\Index{texTstarRepresentation}
   {free \Ts representation of group}%
   {free representation of group}%
\Index{texAffine}
   {Frenet transport}%
   {Frenet transport}%
\Index{texCalculus}
   {function continuous with respect to set of arguments}%
   {function continuous with respect to set of arguments}%
\Index{texCalculus}
   {function of $\gi n$ $D$\Hyph valued variables}%
   {function of n D valued variables}%
\SetIndexSpace%
\Index{texRefernceFrame}
   {$G$\Hyph reference frame}%
   {G reference frame}%
\Index{texTypeBasis}
   {\Gbasis}%
   {G-basis}%
\Index{texBasis}
   {\Gbasis\ of vector space}%
   {G-basis}%
\Index{texBasis}
   {\Gcoords\ of basis}%
   {G-coordinates}%
\Index{texTypeBasis}
   {\Gcoords}%
   {G-coordinates}%
\Index{texTypeBasis}
   {\Gspace}%
   {GSpace}%
\Index{texDrcBasis}
   {geometrical object}%
   {}%
\SubIndex{texDrcBasis}
   {defined in vector space}%
   {geometrical object, vector space}%
\SubIndex{texDrcBasis}
   {in coordinate representation defined in vector space}%
   {geometrical object, coordinate vector space}%
\SubIndex{texDrcBasis}
   {of type $A$}%
   {geometrical object of type A, vector space}%
\Index{texBasis}
   {geometrical object in coordinate representation}%
   {geometrical object, coordinate vector space}%
\Index{texBasis}
   {geometrical object in vector space}%
   {geometrical object, vector space}%
\Index{texBasis}
   {geometrical object of type $A$ in vector space}%
   {geometrical object of type A, vector space}%
\Index{texGroupRing}
   {group algebra}%
   {group algebra}%
\Index{texBasis}
   {\Gspace}%
   {GSpace}%
\SetIndexSpace%
\Index{texBiring}
   {Hadamard inverse of matrix}%
   {Hadamard inverse of matrix}%
\Index{texRefernceFrame}
   {holonomic coordinates of connection}%
   {holonomic coordinates of connection}%
\Index{texRefernceFrame}
   {holonomic coordinates of vector}%
   {vector holonomic coordinates}%
\Index{texFiberedGroup}
   {homogeneous bundle of fibered group}%
   {homogeneous bundle of fibered group}%
\Index{texTstarRepresentation}
   {homogeneous space of group}%
   {homogeneous space of group}%
\Index{texRepresentation}
   {homogeneous space of group}%
   {homogeneous space of group}%
\Index{texFiberedAlgebra}
   {homomorphism of fibered $\mathfrak{F}$\Hyph algebras}%
   {homomorphism of fibered F-algebras}%
\Index{texFiberedGroup}
   {homomorphism of fibered groups}%
   {homomorphism of fibered groups}%
\SetIndexSpace%
\Index{texLieRepresentation}
   {infinitesimal generator}%
   {infinitesimal generator}%
\Index{texLinearLie}
   {infinitesimal generators of group Lie}%
   {infinitesimal generators of group Lie}%
\Index{texDrcBasis}
   {invariance principle}%
   {invariance principle}%
\Index{texBasis}
   {invariance principle in vector space}%
   {invariance principle, vector space}%
\Index{texBundleRelation}
   {inverse fibered correspondence}%
   {inverse fibered correspondence}%
\Index{texBundleRelation}
   {inverse reduced fibered correspondence}%
   {inverse reduced fibered correspondence}%
\Index{texFiberedAlgebra}
   {isomorphism of fibered $\mathfrak{F}$\Hyph algebras}%
   {isomorphism of fibered F-algebras}%
\SetIndexSpace%
\Index{texFiberedGroup}
   {kernel of inefficiency of representation of fibered group}%
   {kernel of inefficiency of representation of fibered group}%
\Index{texRepresentation}
   {kernel of inefficiency of representation of group}%
   {kernel of inefficiency of representation of group}%
\Index{texTstarRepresentation}
   {kernel of inefficiency of \Ts representation of group $G$}%
   {kernel of inefficiency of representation of group}%
\Index{texDiffProperty}
   {Killing equation}%
   {Killing equation}%
\Index{texDiffProperty}
   {Killing equation of second type}%
   {Killing equation second type}%
\Index{texDiffProperty}
   {Killing vector of second type}%
   {Killing vector second type}%
\Index{texBiring}
   {Kronecker symbol}%
   {Kronecker symbol}%
\SetIndexSpace%
\Index{texLie}
   {left invariant vector field}%
   {left invariant vector}%
\Index{texVectorSpace}
   {left module}%
   {left module}%
\Index{texTstarRepresentation}
   {left shift}%
   {left shift}%
\Index{texFiberedGroup}
   {left shift on fibered group}%
   {Tstar shift, fibered group}%
\Index{texRepresentation}
   {left shift on group}%
   {left shift, group}%
\Index{texLie}
   {left structural constant of Lie algebra}%
   {left structural constant of Lie algebra}%
\Index{texVectorSpace}
   {left vector space}%
   {left vector space}%
\Index{texRepresentation}
   {left-side contravariant representation of group}%
   {left-side contravariant representation}%
\Index{texRepresentation}
   {left-side covariant representation of group}%
   {left-side covariant representation}%
\Index{texTstarMorphism}
   {left-side representation of $\mathfrak{F}$\Hyph algebra $A$ in set $M$}%
   {left-side representation of algebra}%
\Index{texFiberedAlgebra}
   {left-side representation of fibered $\mathfrak{F}$\Hyph algebra}%
   {left-side representation of fibered F-algebra}%
\Index{texRepresentation}
   {left-side representation of group}%
   {left-side representation of group}%
\Index{texTstarMorphism}
   {left-side transformation}%
   {left-side transformation}%
\Index{texFiberedAlgebra}
   {left-side transformation on bundle}%
   {left-side transformation of bundle}%
\Index{texLie}
   {Lie algebra of Lie group}%
   {algebra Lie group Lie}%
\SubIndex{texLie}
   {left defined}%
   {left defined algebra Lie}%
\SubIndex{texLie}
   {right defined}%
   {right defined algebra Lie}%
\Index{texDiffProperty}
   {Lie derivative}%
   {Lie derivative}%
\SubIndex{texDiffProperty}
   {of connection}%
   {Lie derivative of connection}%
\SubIndex{texDiffProperty}
   {of metric}%
   {Lie derivative of metric}%
\Index{texLie}
   {Lie group basic operators}%
   {Lie group basic operators}%
\Index{texBundleRelation}
   {lift of correspondence}%
   {lift of correspondence}%
\Index{texBundle}
   {lift of map}%
   {lift of map}%
\Index{texBundleRelation}
   {limit of correspondence with respect to the filter}%
   {limit of correspondence with respect to the filter}%
\Index{texBundleRelation}
   {limit of filter}%
   {limit of filter}%
\Index{texBundleRelation}
   {limit set of filter}%
   {limit set of filter}%
\Index{texRepresentation}
   {linear representation of group}%
   {linear representation of group}%
\Index{texTstarRepresentation}
   {little group}%
   {little group}%
\Index{texRefernceFrame}
   {local reference frame}%
   {local reference frame}%
\Index{texBundle}
   {locally compact at point $p$ space}%
   {locally compact at point space}%
\Index{texBundle}
   {locally compact space}%
   {locally compact space}%
\Index{texRefernceFrame}
   {Lorentz transformation}%
   {Lorentz transformation}%
\SetIndexSpace%
\Index{texPolyvector}
   {$m$\Hyph dimensional parallelepiped}%
   {m dimensional parallelepiped}%
\Index{texPolyvector}
   {$m$\Hyph vector}%
   {m-vector}%
\Index{texDrcReferenceFrame}
   {map of type $G$ on manifold}%
   {map of type G on manifold}%
\Index{texCartesian}
   {mapping space}%
   {mapping space}%
\Index{texDrcMorphism}
   {matrix of \drc linear map}%
   {matrix of drc linear map}%
\Index{texGeomObject}
   {metric-affine manifold}%
   {metric-affine manifold}%
\Index{texFiberedMorphism}
   {morphism of fibered \Ts representations from $\mathcal{F}$ into $\mathcal{G}$}%
   {morphism of fibered representations from f into g}%
\Index{texTstarMorphism}
   {morphism of \Ts representations from $f$ into $g$}%
   {morphism of representations from f into g}%
\Index{texTstarMorphism}
   {morphism of \Ts representations of $\mathfrak{F}$ algebra}%
   {morphism of representations of F algebra}%
\Index{texTstarMorphism}
   {morphism of \Ts representations of $\mathfrak{F}$\Hyph algebra in $\mathfrak{H}$\Hyph algebra}%
   {morphism of representations of F algebra in H algebra}%
\Index{texFiberedMorphism}
   {morphism of \Ts representations of fibered $\mathfrak{F}$\Hyph algebra}%
   {morphism of representations of fibered F algebra}%
\Index{}
   {movement on basis manifold}%
   {movement transformation}%
\SetIndexSpace%
\Index{texPolymodule}
   {$(n)$\hyph vector space}%
   {(n)-vector space}%
\Index{texBundleRelation}
   {$n$\Hyph ary fibered relation}%
   {fibered relation}%
\Index{texGeomObject}
   {nonmetricity}%
   {nonmetricity}%
\Index{texVectorSpace}
   {nonsingular \drc system of linear equations}%
   {nonsingular system of linear equations}%
\Index{texRepresentation}
   {nonsingular \Ts transformation}%
   {nonsingular transformation}%
\Index{texCalculus}
   {norm on \drc vector space}%
   {norm on drc vector space}%
\Index{texCalculus}
   {normed \drc vector space}%
   {normed drc vector space}%
\SetIndexSpace%
\Index{texFiberedAlgebra}
   {operation on bundle}%
   {operation on bundle}%
\Index{texBundleRelation}
   {opposite fibered preordering}%
   {opposite fibered preordering}%
\Index{texFiberedGroup}
   {orbit of representation of fibered group}%
   {orbit of representation of fibered group}%
\Index{texRepresentation}
   {orbit of representation of group}%
   {orbit of representation of group}%
\Index{texTstarRepresentation}
   {orbit of \Ts representation of group}%
   {orbit of representation of group}%
\Index{texBasis}
   {orthonornal basis}%
   {Orthonornal Basis}%
\SetIndexSpace%
\Index{texGeomObject}
   {parallelogram}%
   {parallelogram}%
\Index{texCalculus}
   {partial derivative of mapping $f$ with respect to variable $v^{\gi a}$}%
   {partial derivative of mapping with respect to variable, skew field}%
\Index{texCalculus}
   {partial derivative of mapping $\Vector f$ with respect to variable $v^{\gi a}$}%
   {partial derivative of mapping with respect to variable, drc vector space}%
\Index{texBasis}
   {passive representation}%
   {passive representation}%
\Index{texBasis}
   {passive transformation on basis manifold}%
   {passive transformation}%
\Index{texDrcBasis}
   {passive \Ts representation}%
   {passive representation}%
\Index{texRefernceFrame}
   {pfaffian derivative}%
   {pfaffian derivative}%
\Index{texPolyvector}
   {polyvector}%
   {polyvector}%
\Index{texNewton}
   {potential energy}%
   {potential energy}%
\Index{texDrcBasis}
   {product of geometrical object and constant}%
   {product of geometrical object and constant}%
\Index{texBasis}
   {product of geometrical object and constant in vector space}%
   {product of geometrical object and constant, vector space}%
\Index{texTstarMorphism}
   {product of morphisms of \Ts representations of $\mathfrak{F}$\Hyph algebra}%
   {product of morphisms of representations of F algebra}%
\Index{texBundle}
   {projection of bundle $\mathcal{E}$ along fiber $E$}%
   {projection of bundle along fiber}%
\SetIndexSpace%
\Index{texBasis}
   {quasi affine transformation on basis manifold}%
   {quasi affine transformation}%
\Index{texBasis}
   {quasi movement on basis manifold}%
   {quasi movement}%
\Index{texFiberedMorphism}
   {quotient bundle}%
   {quotient bundle}%
\SetIndexSpace%
\Index{texBiring}
   {\sups row of matrix}%
   {r row}%
\Index{texVectorSpace}
   {\sups rows \drc vector space}%
   {sups rows drc vector space}%
\Index{texBiring}
   {$r$\hyph row of matrix}%
   {r-row}%
\Index{texBiring}
   {\RC inverse element of biring}%
   {rc-inverse element}%
\Index{texVectorSpace}
   {\RC major minor}%
   {RC-major minor}%
\Index{texVectorSpace}
   {\RC matrix group}%
   {rc-matrix group}%
\Index{texVectorSpace}
   {\RC nonsingular matrix}%
   {RC nonsingular matrix}%
\Index{texBiring}
   {\RC power}%
   {rc power}%
\Index{texBiring}
   {\RC product of matrices}%
   {rc-product of matrices}%
\Index{texBiring}
   {\RC quasideterminant}%
   {RC-quasideterminant}%
\Index{texVectorSpace}
   {\RC rank of matrix}%
   {rc-rank of matrix}%
\Index{texVectorSpace}
   {\RC singular matrix}%
   {RC singular matrix}%
\Index{texVectorSpace}
   {\rcd vector space}%
   {rcd vector space}%
\Index{texCartesian}
   {reduced Cartesian product of bundles}%
   {reduced Cartesian product of bundles}%
\Index{texCartesian}
   {reduced Cartesian product of total spaces}%
   {reduced Cartesian product of total spaces}%
\Index{texBundleRelation,texBundleRelation}
   {reduced fibered correspondence from $\mathcal{A}$ to $\mathcal{B}$}%
   {reduced fibered correspondence from A to B}%
\Index{texBundleRelation}
   {reduced fibered correspondence in $\mathcal{A}$}%
   {reduced fibered correspondence in A}%
\Index{texBiring}
   {reducible biring}%
   {reducible biring}%
\Index{texRefernceFrame}
   {reference frame in event space}%
   {reference frame in event space}%
\Index{texDrcReferenceFrame}
   {reference frame manifold}%
   {reference frame manifold}%
\Index{texBundleRelation}
   {reflexive $2$\Hyph ary fibered relation}%
   {reflexive 2 ary fibered relation}%
\Index{texTstarRepresentation}
   {representation of group}%
   {}%
\SubIndex{texTstarRepresentation}
   {contravariant \Ts}%
   {Tstar contravariant representation of group}%
\SubIndex{texTstarRepresentation}
   {covariant \Ts}%
   {Tstar covariant representation of group}%
\SubIndex{texDrcBasis}
   {\drc linear \sT}%
   {linear representation of group}%
\SubIndex{texTstarRepresentation}
   {effective}%
   {effective representation of group}%
\SubIndex{texDrcBasis}
   {\rcd}%
   {rcd linear representation of group}%
\SubIndex{texTstarRepresentation}
   {\sT}%
   {starT representation of group}%
\SubIndex{texTstarRepresentation}
   {\Ts}%
   {Tstar representation of group}%
\Index{texRepresentation}
   {representation of group}%
   {representation of group}%
\Index{texDrcBasis}
   {representative of geometrical object in vector space}%
   {representative of geometrical object, vector space}%
\Index{texBasis}
   {representative of geometrical object in vector space}%
   {representative of geometrical object, vector space}%
\Index{texBundleRelation}
   {restriction of correspondence $\Phi$ to set $C$}%
   {restriction of correspondence}%
\Index{texLie}
   {right invariant vector field}%
   {right invariant vector}%
\Index{texVectorSpace}
   {right module}%
   {right module}%
\Index{texTstarRepresentation}
   {right shift}%
   {right shift}%
\Index{texRepresentation}
   {right shift on group}%
   {right shift, group}%
\Index{texLie}
   {right structural constant of Lie algebra}%
   {right structural constant of Lie algebra}%
\Index{texVectorSpace}
   {right vector space}%
   {right vector space}%
\Index{texRepresentation}
   {right-side contravariant representation of group}%
   {right-side contravariant representation}%
\Index{texRepresentation}
   {right-side covariant representation of group}%
   {right-side covariant representation}%
\Index{texTstarMorphism}
   {right-side representation of $\mathfrak{F}$\Hyph algebra $A$ in set $M$}%
   {right-side representation of algebra}%
\Index{texFiberedAlgebra}
   {right-side representation of fibered $\mathfrak{F}$\Hyph algebra}%
   {right-side representation of fibered F-algebra}%
\Index{texRepresentation}
   {right-side representation of group}%
   {right-side representation of group}%
\Index{texTstarMorphism}
   {right-side transformation}%
   {right-side transformation}%
\Index{texRepresentation}
   {right-side transformation}%
   {right-side transformation}%
\Index{texRepresentation}
   {row vector}%
   {row vector}%
\Index{texVectorSpace}
   {$R\star$\Hyph module}%
   {Rstar-module}%
\SetIndexSpace%
\Index{texNewton}
   {scalar potential}%
   {scalar potential}%
\Index{texNewton}
   {second Newton law}%
   {Second Newton law}%
\Index{texPolyvector}
   {simple polyvector}%
   {simple polyvector}%
\Index{texTstarMorphism}
   {single transitive representation of algebra $A$}%
   {single transitive representation of algebra}%
\Index{texFiberedAlgebra}
   {single transitive representation of fibered $\mathfrak{F}$\Hyph algebra}%
   {single transitive representation of fibered F-algebra}%
\Index{texRepresentation}
   {single transitive representation of group}%
   {single transitive representation of group}%
\Index{texPolyvector}
   {skew product of vectors}%
   {skew product of vectors}%
\Index{texTstarRepresentation}
   {space of orbits of \Ts representation}%
   {space of orbits of Ts representation}%
\Index{texTidal}
   {speed of deviation}%
   {speed of deviation}%
\Index{texDrcMorphism}
   {$(S\RCstar,T\RCstar)$\Hyph linear map of vector spaces}%
   {src trc linear map of vector spaces}%
\Index{texTstarRepresentation}
   {stability group}%
   {stability group}%
\Index{texDrcBasis}
   {standard coordinates of basis}%
   {standard coordinates of basis}%
\Index{texBasis}
   {standard coordinates of basis}%
   {standard coordinates of basis}%
\Index{texBiring}
   {standard representation of matrix}%
   {Standard representation}%
\Index{texVectorSpace}
   {$\star D$\hyph vector space}%
   {starD-vector space}%
\Index{texLinearMap}
   {$\star D$\Hyph product of \drc linear map $A$ over scalar}%
   {starD product of drc linear map over scalar}%
\Index{texVectorSpace}
   {$\star R$\hyph module}%
   {starR-module}%
\Index{texTstarRepresentation}
   {\sT shift}%
   {starT shift}%
\Index{texFiberedGroup}
   {\sT shift on fibered group}%
   {starT shift, fibered group}%
\Index{texTstarMorphism}
   {\sT representation of $\mathfrak{F}$\Hyph algebra $A$ in set $M$}%
   {starT representation of algebra}%
\Index{texFiberedAlgebra}
   {\sT representation of fibered $\mathfrak{F}$\Hyph algebra}%
   {starT representation of fibered F-algebra}%
\Index{texFiberedGroup}
   {\sT representation of fibered group}%
   {starT representation of fibered group}%
\Index{texTstarMorphism}
   {\sT transformation}%
   {starT transformation}%
\Index{texFiberedAlgebra}
   {\sT transformation on bundle}%
   {starT transformation of bundle}%
\SubIndex{}
   {nonsingular}%
   {nonsingular transformation of bundle}%
\Index{texBundle}
   {subbundle}%
   {subbundle}%
\Index{texLinearMap}
   {sum of \drc linear maps}%
   {sum of drc linear maps, drc vector spaces}%
\Index{texDrcBasis}
   {sum of geometrical objects}%
   {sum of geometrical objects}%
\Index{texBasis}
   {sum of geometrical objects in vector space}%
   {sum of geometrical objects, vector space}%
\Index{texBundleRelation}
   {symmetric $2$\Hyph ary fibered relation}%
   {symmetric 2 ary fibered relation}%
\Index{texBasis}
   {symmetry group}%
   {symmetry group}%
\Index{texDrcBasis}
   {symmetry group}%
   {SymmetryGroup}%
\Index{texGenRelativity}
   {synchronization of reference frame}%
   {synchronization of reference frame}%
\SetIndexSpace%
\Index{texLie}
   {tensor product of representations}%
   {tensor product of representations}%
\Index{texCalculus}
   {topological \drc vector space}%
   {topological drc vector space}%
\Index{texCalculus}
   {topological skew field}%
   {topological skew field}%
\Index{texGeomObject}
   {torsion form}%
   {torsion form}%
\Index{texGeomObject}
   {torsion tensor}%
   {torsion tensor}%
\Index{texFiberedMorphism}
   {tower of bundles}%
   {tower of bundles}%
\Index{texTstarMorphism}
   {transformation of set}%
   {transformation of set}%
\Index{texDrcBasis}
   {transformation on basis manifold}%
   {}%
\SubIndex{texDrcBasis}
   {active}%
   {active transformation, vector space}%
\SubIndex{texTypeBasis}
   {affine}%
   {affine transformation}%
\SubIndex{texTypeBasis}
   {movement}%
   {movement transformation}%
\SubIndex{texDrcBasis}
   {passive}%
   {passive transformation, vector space}%
\SubIndex{texTypeBasis}
   {quasi affine}%
   {quasi affine transformation}%
\SubIndex{texTypeBasis}
   {quasi movement}%
   {quasi movement}%
\Index{texFiberedAlgebra}
   {transformation on bundle}%
   {transformation of bundle}%
\Index{texBundleRelation}
   {transitive $2$\Hyph ary fibered relation}%
   {transitive 2 ary fibered relation}%
\Index{texTstarMorphism}
   {transitive representation of $\mathfrak{F}$\Hyph algebra $A$}%
   {transitive representation of algebra}%
\Index{texFiberedAlgebra}
   {transitive representation of fibered $\mathfrak{F}$\Hyph algebra}%
   {transitive representation of fibered F-algebra}%
\Index{texRepresentation}
   {transitive representation of group}%
   {transitive representation of group}%
\Index{texVectorSpace}
   {\Ts linear composition of  vectors}%
   {linear composition of  vectors}%
\Index{texVectorSpace}
   {\Ts matrices vector space}%
   {matrices vector space}%
\Index{texTstarRepresentation}
   {\Ts shift}%
   {Tstar shift}%
\Index{texTstarMorphism}
   {\Ts representation of $\mathfrak{F}$\Hyph algebra $A$ in set $M$}%
   {Tstar representation of algebra}%
\Index{texFiberedAlgebra}
   {\Ts representation of fibered $\mathfrak{F}$\Hyph algebra}%
   {Tstar representation of fibered F-algebra}%
\Index{texFiberedGroup}
   {\Ts representation of fibered group}%
   {Tstar representation of fibered group}%
\Index{texTstarMorphism}
   {\Ts transformation}%
   {Tstar transformation}%
\Index{texFiberedAlgebra}
   {\Ts transformation on bundle}%
   {Tstar transformation of bundle}%
\Index{texFiberedGroup}
   {twin representations of fibered group}%
   {twin representations of fibered group}%
\Index{texTstarRepresentation}
   {twin representations of group}%
   {twin representations of group}%
\Index{texLinearMap}
   {twin representations of skew field}%
   {twin representations of skew field}%
\SetIndexSpace%
\Index{texVectorSpace}
   {unitarity law}%
   {}%
\SubIndex{texVectorSpace}
   {for \Ts vector space}%
   {unitarity law, Tstar vector space}%
\SetIndexSpace%
\Index{texPolymodule}
   {($D_1\RCstar$, ..., $D_n\RCstar$)\hyph vector space}%
   {(d1rc,dnrc)-vector space}%
\Index{texPolymodule}
   {($S\star$, $\star T$)\hyph vector space}%
   {(Sstar,starT)-vector space}%
\Index{texCalculus}
   {valued skew field}%
   {valued skew field}%
\Index{texFiberedAlgebra}
   {vector bundle}%
   {vector bundle}%
\Index{texNewton}
   {vector potential}%
   {vector potential}%
\Index{texVectorSpace}
   {vector space type}%
   {vector space type}%

\CloseIndex

\def\indexname{Special Symbols and Notations}
\OpenIndex

\SetIndexSpace%
\Symb{texBiring}
   {$(^{\gi a}_{\gi b})$\hyph\CR quasideterminant}%
   {a b CR quasideterminant definition}%
\Symb{texBiring}
   {$(^{\gi a}_{\gi b})$\hyph \RC quasideterminant}%
   {a b RC-quasideterminant definition}%
\Symb{texBiring}
   {minor}%
   {A from b a}%
\Symb{texBiring}
   {minor}%
   {A from columns T}%
\Symb{texBiring}
   {minor}%
   {A from rows S}%
\Symb{texBiring}
   {minor}%
   {A without column a}%
\Symb{texBiring}
   {minor}%
   {A without columns T}%
\Symb{texBiring}
   {minor}%
   {A without row b}%
\Symb{texBiring}
   {minor}%
   {A without rows S}%
\Symb{texPolymodule}
   {active transformation}%
   {active transformation}%
\Symb{texTypeBasis}
   {affine space}%
   {affine space}%
\Symb{texBasis}
   {affine space}%
   {An}%
\Symb{texBiring}
   {\subs row ($c$\hyph row) of matrix}%
   {c row}%
\Symb{texBiring}
   {\CR power of element $A$ of biring}%
   {cr power}%
\Symb{texBiring}
   {\CR inverse element of biring}%
   {cr-inverse element}%
\Symb{texBiring}
   {\CR product of matrices}%
   {cr-product of matrices}%
\Symb{texVectorSpace}
   {\dcr vector}%
   {dcr vector}%
\Symb{texLie}
   {derivative of left shift}%
   {derivative of left shift}%
\Symb{texLie}
   {derivative of left shift}%
   {derivative of left shift, 1-Parameter Group}%
\Symb{texLie}
   {derivative of right shift}%
   {derivative of right shift}%
\Symb{texLie}
   {}%
   {derivative of right shift}%
\Symb{texLie}
   {derivative of right shift}%
   {derivative of right shift, 1-Parameter Group}%
\Symb{texLie}
   {derivative of left shift}%
   {derivative of Tstar shift}%
\Symb{texVectorSpace}
   {\drc vector}%
   {drc vector}%
\Symb{texAffine}
   {derivative}%
   {overline nabla_l, definition 2}%
\Symb{texPolymodule}
   {passive transformation}%
   {passive transformation}%
\Symb{texBiring}
   {\sups row ($r$\hyph row) of matrix}%
   {r row}%
\Symb{texBiring}
   {\RC power of element $A$ of biring}%
   {rc power}%
\Symb{texBiring}
   {\RC inverse element of biring}%
   {rc-inverse element}%
\Symb{texBiring}
   {\RC product of matrices}%
   {rc-product of matrices}%
\Symb{texBiring}
   {\RC quasideterminant}%
   {RC-quasideterminant definition}%
\Symb{texTstarRepresentation}
   {right shift}%
   {right shift}%
\Symb{texPolyvector}
   {skew product of vectors $\Vector a_1$, ..., $\Vector a_m$}%
   {skew product of vectors}%
\Symb{texFiberedGroup}
   {\sT shift}%
   {starT shift, fibered group}%
\Symb{texTstarRepresentation}
   {left shift}%
   {Tstar shift}%
\Symb{texFiberedGroup}
   {\Ts shift}%
   {Tstar shift, fibered group}%
\Symb{texRefernceFrame}
   {anholonomic coordinates of vector}%
   {vector anholonomic coordinates}%
\Symb{texRefernceFrame}
   {holonomic coordinates of vector}%
   {vector holonomic coordinates}%

\SetIndexSpace%
\Symb{texBasis}
   {basis manifold of affine space}%
   {BAn}%
\Symb{texBasis}
   {basis manifold of vector space}%
   {basis manifold of vector space}%
\Symb{texBasis}
   {basis manifold of vector space $\mathcal{V}$}%
   {basis manifold of vector space}%
\Symb{texBasis}
   {basis manifold of central affine space}%
   {BCAn}%
\Symb{texBasis}
   {basis manifold of Euclid space}%
   {BEn}%
\Symb{texBundle}
   {Cartesian power $\mathcal{A}$ of bundle $\mathcal{B}$}%
   {Cartesian power of bundle}%
\Symb{texBundle}
   {Cartesian power $A$ of set $B$}%
   {Cartesian power of set}%
\Symb{texTypeBasis}
   {basis manifold of affine space}%
   {FAn}%
\Symb{texTypeBasis}
   {basis manifold of central affine space}%
   {FCAn}%
\Symb{texTypeBasis}
   {basis manifold of Euclid space}%
   {FEn}%

\SetIndexSpace%
\Symb{texBasis}
   {central affine space}%
   {CAn}%
\Symb{texTypeBasis}
   {central affine space}%
   {central affine space}%
\Symb{texLie}
   {left structural constant of Lie algebra}%
   {left structural constant of Lie algebra}%
\Symb{texLie}
   {right structural constant of Lie algebra}%
   {right structural constant of Lie algebra}%

\SetIndexSpace%
\Symb{texLieRepresentation}
   {basis vector of \sT representation}%
   {basis vector of starT representation}%
\Symb{texLieRepresentation}
   {basis vector of \sT representation}%
   {basis vector of starT representation, coordinates}%
\Symb{texLieRepresentation}
   {basis vector of \Ts representation}%
   {basis vector of Tstar representation}%
\Symb{texLieRepresentation}
   {basis vector of \Ts representation}%
   {basis vector of Tstar representation, coordinates}%
\Symb{texVectorSpace}
   {\subs rows \dcr vector space}%
   {c rows dcr vector space}%
\Symb{texCalculus}
   {differential of function}%
   {differential, drc vector space to drc vector space}%
\Symb{texCalculus}
   {differential of function}%
   {differential, drc vector space to skew field}%
\Symb{texVectorSpace}
   {\drc coordinate vector space}%
   {drc coordinate vector space}%
\Symb{texVectorSpace}
   {matrices vector space}%
   {matrices vector space}%
\Symb{texAffine}
   {Cartan derivative}%
   {overbrace D}%
\Symb{texAffine}
   {derivative}%
   {overline D}%
\Symb{texCalculus}
   {partial derivative of mapping $\Vector f$ with respect to variable $v^{\gi a}$}%
   {partial derivative of mapping, 1, drc vector space}%
\Symb{texCalculus}
   {partial derivative of mapping $f$ with respect to variable $v^{\gi a}$}%
   {partial derivative of mapping, 1, skew field}%
\Symb{texVectorSpace}
   {\sups rows \drc vector space}%
   {r rows drc vector space}%
\Symb{texTidal}
   {speed of deviation}%
   {speed of deviation}%
\Symb{texVectorSpace}
   {vector space type}%
   {vector space type}%

\SetIndexSpace%
\Symb{texTypeBasis}
   {affine basis}%
   {Affine Basis}%
\Symb{texBasis}
   {affine basis}%
   {Affine Basis}%
\Symb{texTypeBasis}
   {basis}%
   {basis}%
\Symb{texBasis}
   {basis of vector space}%
   {Basis e}%
\Symb{texBasis}
   {basis in vector space $\mathcal{V}$}%
   {basis in V}%
\Symb{texVectorSpace}
   {basis of vector space}%
   {basis, vector space}%
\Symb{texPolymodule}
   {basis of $(n)$\hyph vector space}%
   {basis,n vector space}%
\Symb{texCartesian}
   {Cartesian power of total spaces}%
   {Cartesian power of total spaces}%
\Symb{texCartesian}
   {Cartesian product of total spaces}%
   {Cartesian product of total spaces, definition 1}%
\Symb{texBasis}
   {central affine basis}%
   {Central Affine Basis}%
\Symb{texRefernceFrame}
   {form of reference frame}%
   {dual forms, reference frame}%
\Symb{texBasis}
   {Euclid space}%
   {En}%
\Symb{texTypeBasis}
   {Euclid space}%
   {En}%
\Symb{texTypeBasis}
   {pseudo Euclid space}%
   {Enm}%
\Symb{texBasis}
   {pseudo Euclid space}%
   {Enm}%
\Symb{texFiberedAlgebra}
   {identical transformation of bundle}%
   {identical transformation of bundle}%
\Symb{texBasis}
   {orthonornal basis}%
   {Orthonornal Basis}%
\Symb{texCartesian}
   {reduced Cartesian product of total spaces}%
   {reduced Cartesian product of total spaces, definition 1}%
\Symb{texFiberedAlgebra}
   {set of nonsingular \sT transformations of bundle $\mathcal{E}$}%
   {set of starT nonsingular transformations of bundle}%
\Symb{texFiberedAlgebra}
   {set of nonsingular \Ts transformations of bundle $\mathcal{E}$}%
   {set of Tstar nonsingular transformations of bundle}%
\Symb{texBasis}
   {standard coordinates of basis}%
   {standard coordinates of basis}%
\Symb{texRefernceFrame}
   {standard coordinates of reference frame}%
   {standard coordinates of reference frame}%
\Symb{texRefernceFrame}
   {vector field of reference frame}%
   {vector field of reference frame}%
\Symb{texBasis}
   {vector of basis}%
   {vector of basis}%

\SetIndexSpace%
\Symb{texVectorSpace}
   {coordinates of basis in \subs rows \dcr vector space}%
   {basis coordinates, c rows dcr vector space}%
\Symb{texVectorSpace}
   {coordinates of basis in \sups rows \drc vector space}%
   {basis coordinates, r rows drc vector space}%
\Symb{texVectorSpace}
   {basis for \subs rows \dcr vector space}%
   {basis, c rows dcr vector space}%
\Symb{texVectorSpace}
   {basis for \sups rows \drc vector space}%
   {basis, r rows drc vector space}%
\Symb{texDiffEq}
   {central affine basis}%
   {Central Affine Basis}%
\Symb{texBundle}
   {fibered morphism from bundle $\mathcal{A}$ into $\mathcal{B}$}%
   {fibered morphism from A into B}%
\Symb{texBundleRelation}
   {filter $\mathfrak{F}$ converges to set $A$}%
   {filter converges}%
\Symb{texFiberedAlgebra}
   {homomorphism of fibered $\mathfrak{F}$\Hyph algebras}%
   {homomorphism of fibered F-algebras}%
\Symb{texBundleRelation}
   {inverse fibered correspondence}%
   {inverse fibered correspondence, 1}%
\Symb{texBundleRelation}
   {inverse reduced fibered correspondence}%
   {inverse reduced fibered correspondence, 1}%
\Symb{texCartesian}
   {map to Cartesian product}%
   {map to Cartesian product}%
\Symb{texTstarRepresentation}
   {representation orbit of group $G$}%
   {orbit of representation of group}%
\Symb{texTypeBasis}
   {orthonornal basis}%
   {Orthonornal Basis}%
\Symb{texRefernceFrame}
   {reference frame}%
   {reference frame}%
\Symb{texRefernceFrame}
   {reference frame, extensive definition}%
   {reference frame, extensive definition}%
\Symb{texPolymodule}
   {standard coordinates of basis}%
   {standard coordinates of basis}%
\Symb{texPolymodule}
   {vector of basis}%
   {vector of basis}%

\SetIndexSpace%
\Symb{texVectorSpace}
   {\CR matrix group}%
   {cr-matrix group}%
\Symb{texFiberedMorphism}
   {fibered little group of section $h$}%
   {fibered little group}%
\Symb{texFiberedMorphism}
   {fibered stability group of section $h$}%
   {fibered stability group}%
\Symb{texLie}
   {algebra Lie of group Lie}%
   {g}%
\Symb{texLie}
   {left defined algebra Lie of group Lie}%
   {gl}%
\Symb{texTypeBasis}
   {affine transformation group}%
   {GLAn}%
\Symb{texBasis}
   {affine transformation group}%
   {GLAn}%
\Symb{texLie}
   {right defined algebra Lie of group Lie}%
   {gr}%
\Symb{texBasis}
   {group of homomorphisms of vector space $\mathcal{V}$}%
   {GV}%
\Symb{texTstarRepresentation}
   {little group of $x$}%
   {little group}%
\Symb{texFiberedGroup}
   {orbit of effective covariant \sT representation of fibered group}%
   {orbit of effective starT covariant representation of fibered group}%
\Symb{texTstarRepresentation}
   {orbit of effective covariant \sT representation of group}%
   {orbit of effective starT covariant representation of group}%
\Symb{texFiberedGroup}
   {orbit of effective covariant \Ts representation of fibered group}%
   {orbit of effective Tstar covariant representation of fibered group}%
\Symb{texTstarRepresentation}
   {orbit of effective covariant \Ts representation of group}%
   {orbit of effective Tstar covariant representation of group}%
\Symb{texVectorSpace}
   {\RC matrix group}%
   {rc-matrix group}%
\Symb{texTstarRepresentation}
   {stability group of $x$}%
   {stability group}%

\SetIndexSpace%
\Symb{texBiring}
   {Hadamard inverse of matrix}%
   {Hadamard inverse of matrix}%
\Symb{texLinearMap}
   {\rcd vector space of \drc linear maps}%
   {rcd vector space of drc linear maps}%

\SetIndexSpace%
\Symb{texLieRepresentation}
   {infinitesimal generator of representation}%
   {infinitesimal generator of representation}%
\Symb{texLinearLie}
   {Lie group infinitesimal generators}%
   {Lie group infinitesimal generators}%

\SetIndexSpace%
\Symb{texRepresentation}
   {left shift}%
   {left shift}%
\Symb{texDiffProperty}
   {Lie derivative of connection}%
   {Lie derivative of connection}%
\Symb{texDiffProperty}
   {Lie derivative of metric}%
   {Lie derivative of metric}%
\Symb{texBundleRelation}
   {limit of correspondence $\Phi$ with respect to the filter $\mathfrak{F}$}%
   {limit of correspondence with respect to the filter}%
\Symb{texBasis}
   {passive transformation}%
   {passive transformation}%
\Symb{texRepresentation}
   {set of left-side nonsingular transformations of set $M$}%
   {set of left-side nonsingular transformations}%

\SetIndexSpace%
\Symb{texTstarMorphism}
   {set of \sT transformations of set $M$}%
   {set of starT transformations}%
\Symb{texTstarMorphism}
   {set of \Ts transformations of set $M$}%
   {set of Tstar transformations}%
\Symb{texTstarRepresentation}
   {space of orbits of effective \sT covariant representation of the group}%
   {space of orbits of effective sT representation}%
\Symb{texTstarRepresentation}
   {space of orbits of effective \Ts covariant representation of the group}%
   {space of orbits of effective Ts representation}%
\Symb{texTstarRepresentation}
   {space of orbits of \Ts representation $f$ of group $G$ in set $M$}%
   {space of orbits of Ts representation}%

\SetIndexSpace%
\Symb{texBasis}
   {geometrical object in coordinate representation}%
   {geometrical object, coordinate vector space}%
\Symb{texBasis}
   {geometrical object}%
   {geometrical object, vector space}%
\Symb{texFiberedGroup}
   {orbit of representation of fibered group $\mathcal{G}$}%
   {orbit of representation of fibered group}%
\Symb{texRepresentation}
   {orbit of representation of the group $G$}%
   {orbit of representation of group}%

\SetIndexSpace%
\Symb{texBundle}
   {bundle}%
   {bundle}%
\Symb{texFiberedMorphism}
   {bundle of level $2$}%
   {bundle of level 2}%
\Symb{texFiberedMorphism}
   {bundle of level $n$}%
   {bundle of level n}%
\Symb{texCartesian}
   {Cartesian power of bundle}%
   {Cartesian power of bundle}%
\Symb{texCartesian}
   {Cartesian product of bundles}%
   {Cartesian product of bundles, definition 1}%
\Symb{texCartesian}
   {reduced Cartesian product of bundles}%
   {reduced Cartesian product of bundles, definition 1}%
\Symb{texFiberedAlgebra}
   {set of nonsingular \sT transformations of bundle $\bundle{}pE{}$}%
   {set of starT nonsingular transformations of bundle, projection}%
\Symb{texFiberedAlgebra}
   {set of nonsingular \Ts transformations of bundle $\bundle{}pE{}$}%
   {set of Tstar nonsingular transformations of bundle, projection}%

\SetIndexSpace%
\Symb{texBasis}
   {active transformation}%
   {active transformation}%
\Symb{texAffine}
   {Cartan curvature}%
   {Cartan curvature}%
\Symb{texVectorSpace}
   {\CR rank of matrix}%
   {cr-rank of matrix}%
\Symb{texBundleRelation}
   {diagonal in bundle  $\bundle{}pA{}$}%
   {diagonal in bundle, 2}%
\Symb{texBundleRelation}
   {diagonal in bundle $\mathcal{A}$}%
   {diagonal in reduced bundle, 2}%
\Symb{texAffine}
   {curvature}%
   {GLn curvature_overline}%
\Symb{texVectorSpace}
   {\RC rank of matrix}%
   {rc-rank of matrix}%
\Symb{texRepresentation}
   {right shift}%
   {right shift}%
\Symb{texRepresentation}
   {set of right-side nonsingular transformations of set $M$}%
   {set of right-side nonsingular transformations}%

\SetIndexSpace%
\Symb{texBundleRelation}
   {composition of fibered correspondences}%
   {composition of fibered correspondences}%
\Symb{texBundleRelation}
   {inverse fibered correspondence}%
   {inverse fibered correspondence, 2}%
\Symb{texBundleRelation}
   {inverse reduced fibered correspondence}%
   {inverse reduced fibered correspondence, 2}%
\Symb{texVectorSpace}
   {linear span in vector space}%
   {linear span, vector space}%

\SetIndexSpace%
\Symb{texLie}
   {tangent plane to group $G$}%
   {TaG}%

\SetIndexSpace%
\Symb{texBasis}
   {coordinate vector space}%
   {coordinate vector space}%
\Symb{texBasis}
   {coordinates in vector space}%
   {coordinates in vector space}%
\Symb{texVectorSpace}
   {\dcr vector space}%
   {left CR vector space}%
\Symb{texVectorSpace}
   {\drc vector space}%
   {left RC vector space}%
\Symb{texLinearMap}
   {($S$, $T$)\hyph bimodule}%
   {R S bimodule}%
\Symb{texVectorSpace}
   {\crd vector space}%
   {right CR vector space}%
\Symb{texVectorSpace}
   {\rcd vector space}%
   {right RC vector space}%
\Symb{texBasis}
   {vector space}%
   {V}%

\SetIndexSpace%
\Symb{texPolymodule}
   {geometrical object in coordinate representation		defined in vector space}%
   {geometrical object, coordinate vector space}%
\Symb{texPolymodule}
   {geometrical object in vector space}%
   {geometrical object, vector space}%

\SetIndexSpace%
\Symb{texRefernceFrame}
   {anholonomic coordinate}%
   {x(k)}%

\SetIndexSpace%
\Symb{texBundleRelation}
   {diagonal in bundle $\mathcal{A}$}%
   {diagonal in bundle, 1}%

\SetIndexSpace%
\Symb{texTidal}
   {deviation of trajectories}%
   {deviation of trajectories}%
\Symb{texRepresentation}
   {identical transformation}%
   {identical transformation}%
\Symb{texTstarMorphism}
   {identical transformation}%
   {identical transformation}%
\Symb{texBasis}
   {image of vector $\Vector e_k\in\Basis e$ under isomorphism to coordinate vector space}%
   {image of vector e_k, coordinate vector space}%
\Symb{texBiring}
   {Kronecker symbol}%
   {Kronecker symbol}%

\SetIndexSpace%
\Symb{texRefernceFrame}
   {anholonomic coordinates of connection}%
   {anholonomic coordinates of connection}%
\Symb{texAffine}
   {Cartan symbol}%
   {Cartan symbol}%
\Symb{texAffine}
   {connection}%
   {conection overline}%
\Symb{texRefernceFrame}
   {holonomic coordinates of connection}%
   {holonomic coordinates of connection}%
\Symb{texAffine}
   {Cartan connection}%
   {overbrace Gamma i kl}%
\Symb{texBundle}
   {set of sections of bundle}%
   {set of sections of bundle}%

\SetIndexSpace%
\Symb{texLie}
   {inverse operator to operator $\psi_l$}%
   {inverse operator to operator psi l}%
\Symb{texLie}
   {inverse operator to operator $\psi_r$}%
   {inverse operator to operator psi r}%

\SetIndexSpace%
\Symb{texRefernceFrame}
   {anholonomity object}%
   {anholonomity object}%

\SetIndexSpace%
\Symb{texLie}
   {basic operator of group Lie}%
   {Lie Basic Operator L}%
\Symb{texLie}
   {}%
   {Lie Basic Operator L}%
\Symb{texLie}
   {basic operator of group Lie}%
   {Lie Basic Operator L, 1-Parameter Group}%
\Symb{texLie}
   {basic operator of group Lie}%
   {Lie Basic Operator R}%
\Symb{texLie}
   {}%
   {Lie Basic Operator R}%
\Symb{texLie}
   {basic operator of group Lie}%
   {Lie Basic Operator R, 1-Parameter Group}%

\SetIndexSpace%
\Symb{texRefernceFrame}
   {coordinate reference frame}%
   {coordinate reference frame, extensive definition}%
\Symb{texCalculus}
   {partial derivative of mapping $\Vector f$ with respect to variable $v^{\gi a}$}%
   {partial derivative of mapping, 2, drc vector space}%
\Symb{texCalculus}
   {partial derivative of mapping $f$ with respect to variable $v^{\gi a}$}%
   {partial derivative of mapping, 2, skew field}%
\Symb{texRefernceFrame}
   {derivative $e_{(k)}$}%
   {partial(k)}%

\SetIndexSpace%
\Symb{texLie}
   {Lie group composition law}%
   {Lie group composition law}%

\SetIndexSpace%
\Symb{texAffine}
   {Cartan derivative}%
   {overbrace nabla_l}%
\Symb{texAffine}
   {derivative}%
   {overline nabla_l, definition 1}%

\SetIndexSpace%
\Symb{texBundleRelation}
   {restriction of correspondence $\Phi$ to set $C$}%
   {restriction of correspondence}%

\SetIndexSpace%
\Symb{texCartesian}
   {Cartesian product of bundles}%
   {Cartesian product of bundles, definition 2}%
\Symb{texCartesian}
   {Cartesian product of total spaces}%
   {Cartesian product of total spaces, definition 2}%
\Symb{texCartesian}
   {reduced Cartesian product of bundles}%
   {reduced Cartesian product of bundles, definition 2}%
\Symb{texCartesian}
   {reduced Cartesian product of total spaces}%
   {reduced Cartesian product of total spaces, definition 2}%

\SetIndexSpace%
\Symb{texBundle}
   {fibered subset}%
   {fibered subset}%
\Symb{texBundle}
   {subbundle}%
   {subbundle}%

\CloseIndex

\end{document}